\documentclass{aa}
\usepackage{epsfig}
\usepackage{lscape}
\usepackage{aalongtable}
\usepackage{natbib}
\usepackage{graphics}
\usepackage{txfonts}
\usepackage{color}
\usepackage{times}
\newfont{\nsf}{cmssdc10 scaled 1000}

\newcommand{\kmsec}{km s$^{-1}$}
\newcommand{\msun}{$M_\odot$}
\newcommand{\zsun}{$Z_\odot$}

\newcommand{\sbb}{mag/$\sq\arcsec$}
\newcommand{\oh}{12+log(O/H)}

\def\ha{H$\alpha$}
\def\hb{H$\beta$}
\def\o5007{[O {\sc iii}] $\lambda$5007}
\def\h2{H{\sc ii}}
\def\P25{R$_{\rm SF}$}
\def\R25{R$_{\rm host}$}

\newcommand{\PutWin}[4]{
\put(#1,#2){\parbox{#3}{#4}}}
\def\lowspace{\rule[-1.25ex]{0cm}{1.25ex}}
\def\upperspace{\rule[0.0ex]{0cm}{2.5ex}}
\def\rr{{\sl R}$^{\star}$}
\def\P25{{\sl R}$_{\rm SF}$}
\def\E25{{\sl R}$_{\rm host}$}
\def\eqan{\begin{equation}}
\def\eqen{\end{equation}}
\newfont{\lvss}{cmssdc10 scaled 900}
\newfont{\lx}{cmssdc10 scaled 760}
\def\sfha{\lvss SFH1\rm}
\def\sfhb{\lvss SFH2\rm}

\begin{document}
\title{Extremely metal-poor star-forming galaxies
\thanks{Based on observations collected at the European Southern Observatory, 
Chile, ESO programs 075.B-0768 and 076.B-0739.}
}
\subtitle{New detections and general morphological and photometric properties}

\author{P.\ Papaderos \inst{1,2}
\and N. G.\ Guseva \inst{3}
\and Y. I.\ Izotov \inst{3}
\and K. J.\ Fricke \inst{4}}
\offprints{P. Papaderos, papaderos@iaa.es}
\institute{     Instituto de Astrof\'{\i}sica de Andaluc\'{\i}a (CSIC),
  Camino Bajo de Hu\'etor 50, Granada E-18008, Spain
\and
                Max-Planck-Institute for Radioastronomy, Auf dem H\"ugel 69,
                53121 Bonn, Germany
\and
                 Main Astronomical Observatory, 
                 Ukrainian National Academy of Sciences,
                 27 Zabolotnoho str., Kyiv 03680,  Ukraine
\and
Institute for Astrophysics, University of G\"ottingen, Friedrich-Hund-Platz 1, 
37077 G\"ottingen, Germany
}

\date{Received \hskip 2cm; Accepted}

\abstract
% Context
{Extremely metal-deficient [\oh$\la$7.6] emission-line galaxies in the nearby
universe are invaluable laboratories of extragalactic astronomy and 
observational cosmology since they allow us to study collective star formation 
and the evolution of galaxies under chemical conditions
approaching those in distant protogalactic systems. 
However, despite intensive searches over the last three decades, 
nearby star-forming (SF) galaxies with strongly subsolar metallicity 
remain extremely scarce.
}
% Aims
{We searched the Sloan Digital Sky Survey
(SDSS) and the Six-Degree Field Galaxy Redshift Survey (6dFGRS) for promising
low-metallicity candidates using a variety of spectroscopic criteria.
}
% Methods
{We present long-slit spectroscopy with the 3.6m ESO telescope
of eight \h2\ regions in seven emission-line dwarf galaxies, selected from 
the Data Release 4 of SDSS (six galaxies) and from 6dFGRS (one galaxy). 
In addition, we use SDSS imaging data to investigate the photometric structure 
of the sample galaxies.
}
% Results
{From the 3.6m telescope spectra, we determine the oxygen abundance 
of these systems to be \oh$\la$7.6, placing them among the most
metal-poor star-forming galaxies ever discovered.
% -
Our photometric analysis reveals a moderately blue, stellar host galaxy 
in all sample galaxies.}  
% Conclusions
{The detection of a stellar host in all galaxies studied here and all
previously studied extremely metal-deficient SF galaxies implies that 
they are unlikely to be forming their first generation of stars.
With regard to the structural properties of their host galaxy, we demonstrate
that these systems are indistinguishable from blue compact dwarf (BCD) galaxies.
However, in contrast to the majority ($>$90\%) of BCDs that are characterised
by red elliptical host galaxies, extremely metal-poor SF dwarfs (hereafter XBCDs)
reveal moderately blue and irregular hosts.
This is consistent with a young evolutionary status and 
in the framework of standard star formation histories implies
that several XBCDs formed most of their stellar
mass in the past $\sim$2 Gyr. 
A large fraction of XBCDs reveal a \emph{cometary} morphology 
due to the presence of intense SF activity at 
one edge of an elongated host galaxy with a gradually 
decreasing surface brightness towards its antipodal end.
}

\maketitle

\color{black}
% ================================================================ Section 1
\section{Introduction \label{intro}}
% ================================================================ 
The identification and detailed studies of chemically unevolved, 
star-forming (SF) galaxies in the nearby universe, of almost 
pristine chemical composition, is a major task for contemporary 
observational cosmology.  
Some important aspects of these studies are the following.

First, systematic studies of SF galaxies with strongly subsolar 
metal abundances are indispensable for placing tight observational 
constraints on the primordial $^4$He abundance $Y_{\rm p}$ 
\citep[][and references therein]{ITS2007,Peimbert2007}.

Secondly, the metallicity plays a key role in virtually all aspects of 
star- and galaxy evolution since it influences e.g. 
a) the production rate of Lyman continuum photons and 
efficiency of radiative winds in massive stars
\citep[see e.g.][]{L92,SchaererdeKoter1997,Kudritzki2002},
including the properties of Wolf-Rayet stellar populations 
\citep{Izotov1997-IZw18,ShaererVacca1998,Guseva2000-WR,Crowther2006,Brinchman2008}, 
as well as 
b) the cooling efficiency of the hot ($\sim 10^7$ K) X-ray emitting 
gas \citep{BoehringerHensler1989}, recognised to be an 
ubiquitous component of starburst galaxies 
\citep[see e.g.][]{Heckman1995-NGC1569,PapaderosFricke1998a,Martin2002-NGC1569,Ott2005b}.
Spatially resolved studies of extremely metal-poor nearby SF galaxies may therefore
yield crucial insights into the early evolution of faint protogalactic systems
forming out of primordial or almost metal-free gas in the young universe. 

The most suitable nearby objects for exploring these 
issues are blue compact dwarf (BCD) galaxies. 
These galaxies form a morphologically heterogeneous class 
of intrinsically faint ($M_B$ $\ga$ --18 mag) extragalactic systems 
undergoing intense starburst activity on a spatial scale of
typically $\sim$1 kpc 
\citep[see][hereafter P02 and references therein]{Papaderos2002-IZw18}.
In $\sim$90\% of the local BCD population, SF activity proceeds in one or several
luminous \h2\ regions within the central part of a more extended,
low-surface brightness (LSB) elliptical host galaxy 
\citep[][hereafter LT86 and P96a, respectively]{LooseThuan1986,Papaderos1996a}.
This red (0.8$\la B-R\,\,{\rm (mag)}\la$1.2; P96a) galaxy host 
contains, on average, one half of the optical emission
of a BCD \cite[][hereafter P96b]{Papaderos1996b} and typically dominates 
the line-of-sight intensity of this system for surface brightness levels 
fainter than 24.5 $B$ \sbb\ (P96a, P02).
Deep surface photometry in the optical and near infrared
\citep{Noeske2003-NIR,Cairos2003-NIR,GildePazMadore2005}
and colour-magnitude-diagram analyses
\citep[e.g.,][]{SchulteLadbeck1999-7Zw403,Tosi2001-NGC1705} 
confirmed the earlier conclusion (LT86, P96b) that BCDs are,
overwhelmingly, evolved gas-rich dwarfs undergoing 
recurrent starburst activity. 

BCDs are the most metal-deficient emission-line galaxies known in the 
nearby universe.
However, while all BCDs show subsolar chemical abundances, 
it is notoriously difficult to find extremely metal-deficient systems
in the range \oh$\la$7.6. 
The BCD abundance distribution peaks at \oh$\approx$8.1
with a sharp drop-off at lower values
\citep{Terlevich1991,Thuan1995,IT98a,KunthOstlin2000}.
One of the first BCDs discovered, I\,Zw\,18 \citep{SargentSearle1970} 
with \oh=7.17$\pm$0.01 \citep{Izotov1997-IZw18},
held the record as the most metal-deficient SF galaxy known for more than three decades.
Only very recently was this system replaced in the metallicity
ranking by the BCD SBS 0335--052\,W with an oxygen abundance 
\oh=7.12$\pm$0.03 \citep{Izotov2005-SBS0335W}.

Despite large observational efforts over the past three decades and 
systematic studies of several $10^5$ catalogued emission-line galaxies, 
less than 20 BCDs with \oh$\la$7.6 (hereafter XBCDs) were identified 
in the nearby ($z\la 0.04$) universe until a few years ago 
\citep[][and references therein]{KunthOstlin2000}.
Since then substantial progress has been achieved, and more than one dozen further XBCDs 
discovered \citep{K03,K04b,Guseva2003-SBS1129,Guseva2003-HS1442,
Pustilnik2005-DDO68,Pisano2005,Pustilnik2006-HS2134,
Izotov2006-2XBCD,Papaderos2006-2dF,IzotovThuan2007-MMT,Kewley2007-SDSS0909}.
At higher redshift ($0.2 \la z \la 0.8$), dedicated surveys utilising 
10m-class telescopes and optimised search strategies led to the discovery of 
about ten more SF dwarf galaxies with \oh$\la$7.6 and an absolute 
rest-frame magnitude in the range of BCDs \citep{Kakazu2007}.
In spite of the extreme scarcity and intrinsic faintness of XBCDs, 
there is therefore a tangible prospect of unveiling a significant number of 
these systems in the years to come.

\begin{table*}
\caption{Coordinates of the sample galaxies (J2000.0) \label{tab1}}
\begin{tabular}{lccccccc} \hline
Name       & R.A.   & DEC  & redshift & Distance$^{\rm a}$ (Mpc) & airmass & P.A. ($^{\circ}$) & exp.time (sec)\\ 
\hline
J 0133$+$1342   & 01 33 52.6 & $+13$ 42 09  & 0.00879 & 37.8 & 1.37 & 170.4  & 3600 (3 $\times$ 1200) \\ 
g0405204-364859 & 04 05 20.4 & $-36$ 48 59  & 0.00268 & 11.4 & 1.04 & 122.0  & 2400 (3 $\times$ 800) \\
J 1044+0353a    & 10 44 57.8 & $+03$ 53 13  & 0.01274 & 51.2 & 1.19 & 101.5  & 3600 (3 $\times$ 1200) \\
J 1044+0353b    & 10 44 58.0 & $+03$ 53 13  & 0.01284 & 51.2 & 1.19 & 101.5  & 3600 (3 $\times$ 1200) \\
J 1201+0211     & 12 01 22.3 & $+02$ 11 08  & 0.00327 & 14.0 & 1.21 & 119.0  & 2400 (2 $\times$ 1200) \\
J 1414-0208     & 14 14 54.1 & $-02$ 08 23  & 0.00528 & 23.0 & 1.16 & 155.3  & 3600 (3 $\times$ 1200) \\
J 2230-0006     & 22 30 36.8 & $-00$ 06 37  & 0.00559 & 24.9 & 1.33 & 156.0 & 3600 (3 $\times$ 1200) \\
J 2302+0049     & 23 02 10.0 & $+00$ 49 39  & 0.03302 & 134.6 & 1.17 & 156.0 & 2400 (2 $\times$ 1200) \\ \hline
\end{tabular}

\vspace*{0.5ex}\parbox{16cm}{
$^{\rm a}$: Distance, derived after correction of the measured redshifts 
for the motion relative to the Local Group centroid and the Virgocentric flow,
assuming a Hubble constant of 75 km s$^{-1}$ Mpc$^{-1}$.}
\end{table*}

\smallskip
Third, recent work provides strong observational support to the idea that some
XBCDs in the nearby universe formed most of their stellar mass in the past 
$\sim$1 Gyr, which implies that they are cosmologically young systems
\citep{Papaderos1998-SBS0335,Vanzi2000-SBS0335E,Guseva2001-SBS0940,Papaderos2002-IZw18,
Guseva2003-SBS1129,Guseva2003-HS1442,Guseva2003-SBS1415,Hunt2003-IZw18,Pustilnik2004-SBS0335}.
The youth hypothesis for XBCDs is further supported by a number of evolutionary synthesis and
colour-magnitude diagram (CMD) studies that indicate an upper age of 0.1-2 Gyr
for the stellar component of some of those systems
\citep{Izotov1997-SBS0335E,Thuan1997-SBS0335E,Izotov2001-IZw18,
Fricke2001-T1214,IzotovThuan2004-IZw18,IzotovThuan2004-UGC4483}.\\
In the case of the XBCD I Zw 18, the conclusion that this system started
forming stars not earlier than $\la$0.5 Gyr ago \citep{IzotovThuan2004-IZw18}, 
was recently disputed by the subsequent CMD studies of \citet{Aloisi2007-IZw18}.

However, apart from the question of when the first stars in a XBCD were formed,
there appears to be broad consensus that many of these systems
underwent the dominant phase of their formation only recently.
If so, XBCDs may be regarded as convenient laboratories to explore
the main processes driving dwarf galaxy formation, as long as their
morphological and dynamical relics have not had time to be erased in the course of
the secular galactic evolution, through e.g. two-body relaxation,
galaxy merging and subsequent star formation episodes. 
Low-mass protogalaxies in the distant universe, once detected, 
will remain due to the cosmological dimming and their intrinsic 
faintness and compactness challenging to study with sufficient resolution 
and accuracy, even with the next generation of extremely large telescopes.
Young XBCD candidates provide in this respect a bridge between
near-field and high-redshift observational cosmology and  
invaluable laboratories of extragalactic research.

This paper investigates the spectroscopic, photometric, and morphological
properties of a new sample of XBCDs. 
It is organised as follows: the sample selection, data acquisition, 
and reduction are described in Sect. \ref{obs}. 
Our spectroscopic and photometric analysis is presented 
in Sect. \ref{results}, and in Sect. \ref{objects}, we discuss the properties 
of individual sample galaxies. The photometric and morphological properties of XBCDs are reviewed 
in Sect. \ref{discussion}, and in Sect. \ref{summary}, we summarise our conclusions. 

% =====================================================
\section{Sample and observations \label{obs}}
% =====================================================
% ==========================================
\subsection{Sample selection \label{sample}}
% ==========================================
%
% ******************************************************************
%                 F I G U R E  1: Slit Position 
% ******************************************************************
\begin{figure*}[th]
\label{fig:slit}
\begin{picture}(17,19)
\put(0.5,7.4){{\psfig{figure=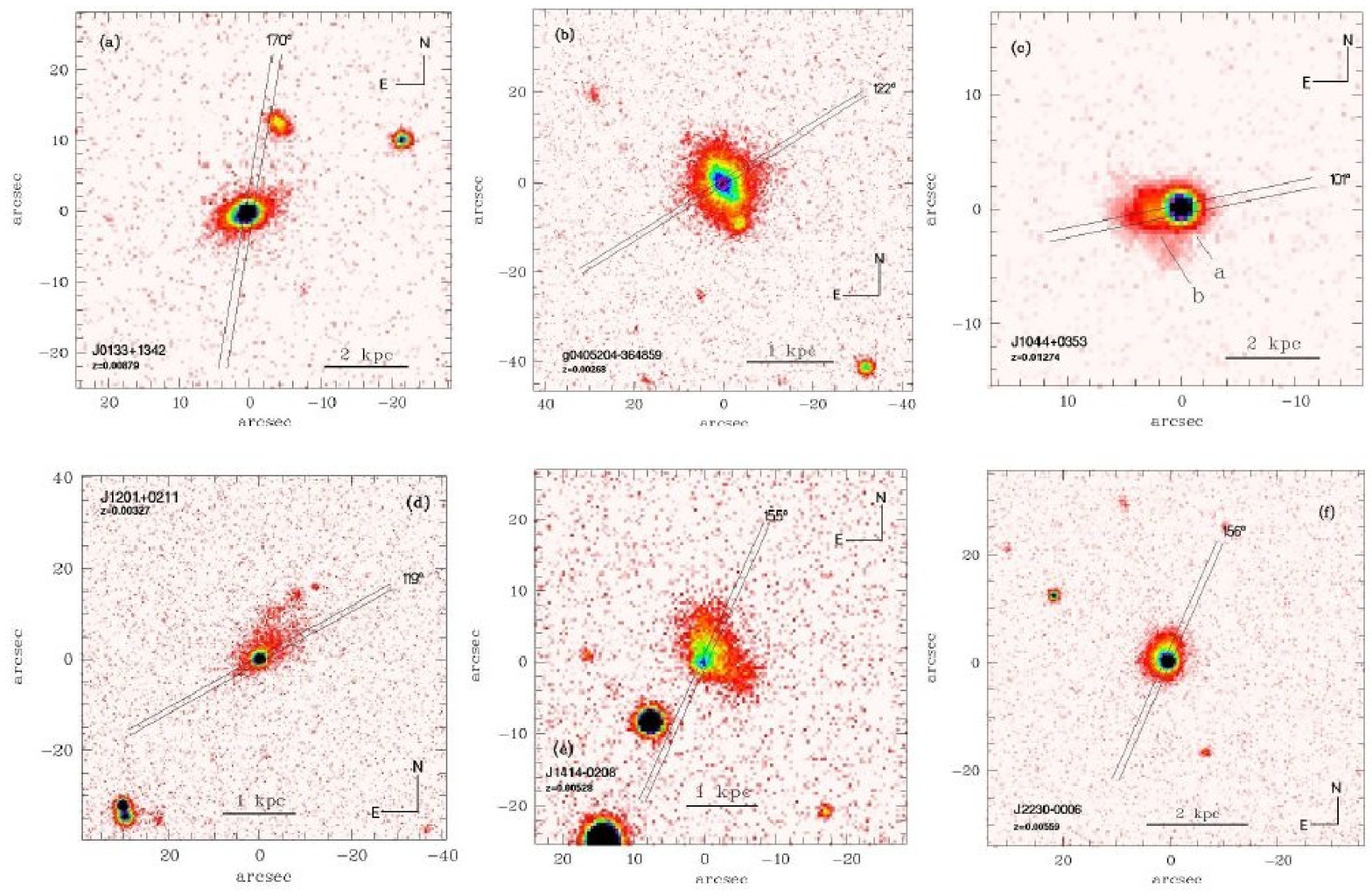,width=17cm,angle=0.,clip=}}}
\put(0.5,0.){{\psfig{figure=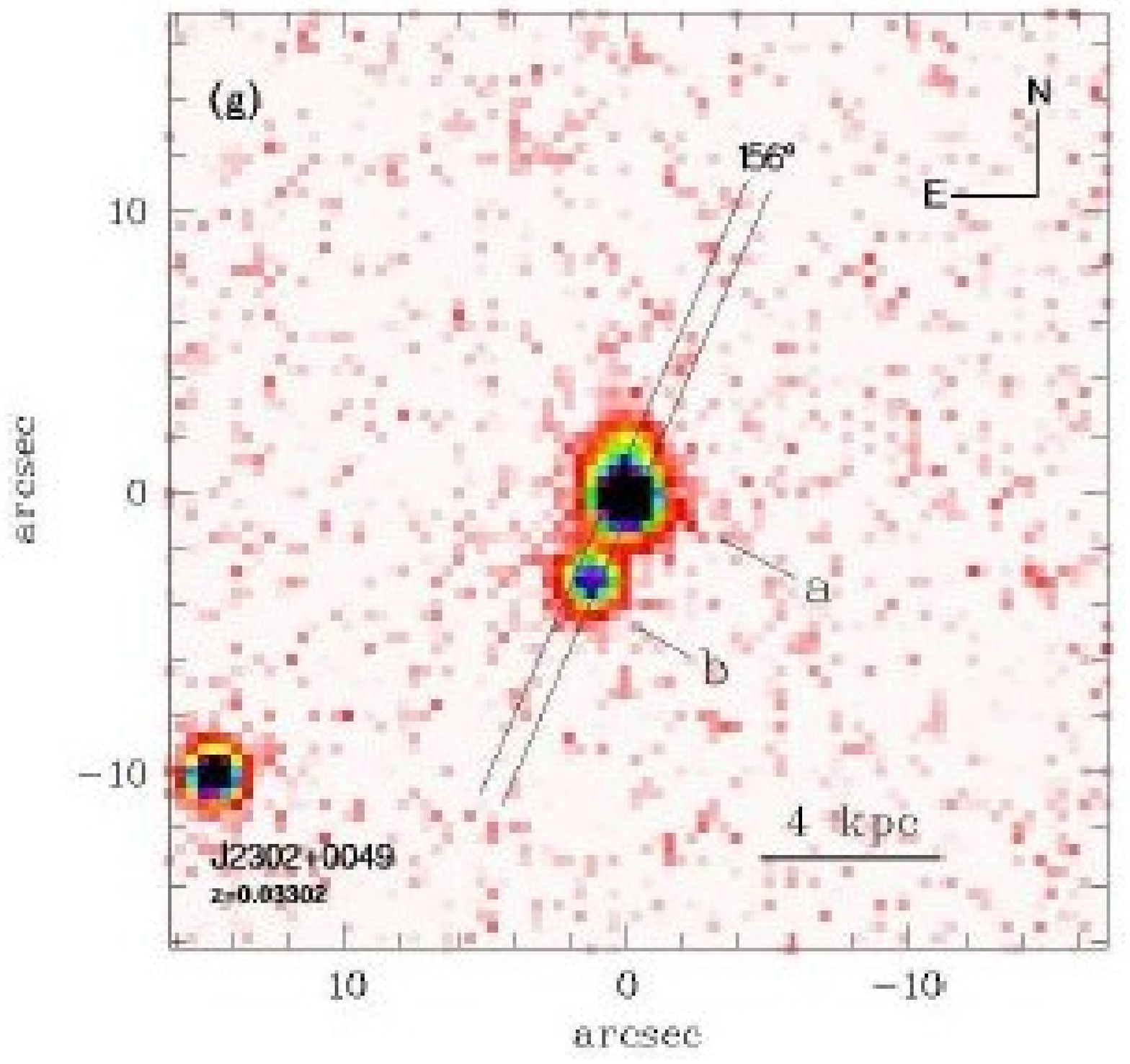,height=5.5cm,angle=0.,clip=}}} 
\PutWin{8}{4}{10cm}{
\caption{Reproduction of the slit position on optical exposures of the sample
 galaxies. The displayed images were obtained through stacking of SDSS $gri$
images (SDSS galaxies) or coaddition of acquisition exposures
 (g0405204-364859, panel b). North is up and east to the left.}}
\end{picture}
\end{figure*}

With the aim of finding new XBCD candidates, we carried out a systematic 
search for these objects in the SDSS Data Release 4 \citep{Adelman2006}, 
on the basis of the relative fluxes of emission lines. 
All extremely metal-deficient emission-line galaxies 
known are characterised by relatively weak (compared to 
H$\beta$) [O~{\sc ii}] $\lambda$3727, 
[O {\sc iii}] $\lambda$4959,$\lambda$5007 and [N {\sc ii}] $\lambda$6583 emission 
lines \citep[e.g.][]{IT98a,IT98b,TI05,Pustilnik2005-DDO68}.
These spectral properties identify uniquely low-metallicity 
dwarfs, since no other type of galaxy 
possess these spectral characteristics.
We considered additionally spectra in which the [O {\sc iii}] $\lambda$4363
emission line is weak or not detected. 
Since the [O {\sc ii}] $\lambda$3727 line is outside of the observed wavelength 
range of the SDSS spectra for all galaxies of redshift $z<0.02$, we used the
[O {\sc iii}] $\lambda$4959, $\lambda$5007, and [N {\sc ii}] $\lambda$6583 
emission to identify extremely low-metallicity galaxy candidates. 
Following this strategy \citep{Izotov2006-2XBCD}, we selected only galaxies with 
weak [O {\sc iii}] and [N {\sc ii}] emission lines. 

We also searched for XBCD candidates in
the Six-Degree Field Galaxy
Redshift Survey (6dFGRS) \citep{Jones2005-6dF}.
Our strategy \citep[see also][hereafter P06]{Papaderos2006-2dF} 
was also based on the visual selection of spectra
with blue continua and strong emission lines, and for which 
the [O {\sc iii}] $\lambda$4363 emission line was present.
This is because the [O {\sc iii}] $\lambda$4363 auroral line is easily detectable 
in high-temperature and hence low-metallicity objects. 
This criterion also selects high-metallicity AGNs, which we rejected by visual
examination of each spectrum based on the relative intensities
of some emission lines (e.g. strong [O {\sc ii}] $\lambda$3727 and
[N {\sc ii}] $\lambda$6583 emission in combination with strong
[O {\sc iii}] $\lambda$5007 emission, strong He {\sc ii} $\lambda$4686,
and [O {\sc i}] $\lambda$6300, etc.). 

We describe observations of seven XBCD candidates
with the 3.6m ESO telescope.
Three candidates were reported to be extremely metal-poor galaxies 
by \citet{K03,K04b} on the basis of oxygen abundances obtained  
from SDSS spectral measurements.
However, given that in the SDSS spectra of these three systems, 
the [O {\sc ii}] $\lambda$3727 line is out of the range, 
the value of O$^+$/H$^+$ was calculated from the 
[O {\sc ii}] $\lambda$7320,7330 lines.
But since the intensities of these infrared lines are many times lower 
than that of the [O {\sc ii}] $\lambda$3727 doublet, this method is applicable
to bright SDSS galaxies only and may be affected by large uncertainties 
compared with the standard method.
In addition, some of the strongest emission lines in SDSS spectra 
are truncated or saturated.
Moreover, the fibers used in the acquisition of SDSS spectra encompass a region of
3\arcsec\ in diameter and are typically positioned on the galaxy center.
As a result, the SDSS data rarely enable spectroscopic studies of off-center SF regions.
For this reason, follow-up spectroscopy of suspected 
XBCD candidates with off-center \h2\ regions is much needed.
This was demonstrated in e.g. \citet{Izotov2006-2XBCD},
who discovered the XBCD \object{J2104-0035} by follow-up 
long-slit spectroscopy of selected SDSS galaxies. 
Star-forming activities in this system proceed mainly in a 
detached metal-poor [\oh=7.26$\pm$0.03] \h2\ region 
$\sim$4\arcsec\ northeast of the geometrical center of the galaxy host, 
which was not included in the spectroscopic data set of SDSS.

The spectroscopic analysis presented here is supplemented by surface photometry. 
This is a step towards the systematization of the photometric and
morphological properties of extremely metal-poor emission-line galaxies, 
which is one of the main goals of our studies.

% ===========================================================
\subsection{Data acquisition and reduction \label{reduction}}
% ===========================================================

New spectroscopic observations of seven \h2\ regions in six 
SDSS galaxies, and one 6dFGRS galaxy (\object{g0405204-364859}) 
were obtained on 10 -- 13 April, 
and 6--8 October, 2005 with the  EFOSC2 (ESO Faint Object Spectrograph 
and Camera) mounted at the 3.6m ESO telescope at La Silla. 
The coordinates of the targets are listed in Table~1.
%%%%%%%%%%%%%%%%%%%%%%%%%%%%%%%%%%%%%%%%%%%%%%%%
%    F I G U R E  2: Spectra
%%%%%%%%%%%%%%%%%%%%%%%%%%%%%%%%%%%%%%%%%%%%%%%%
\begin{figure*}[t]
\hspace*{-0.0cm}\psfig{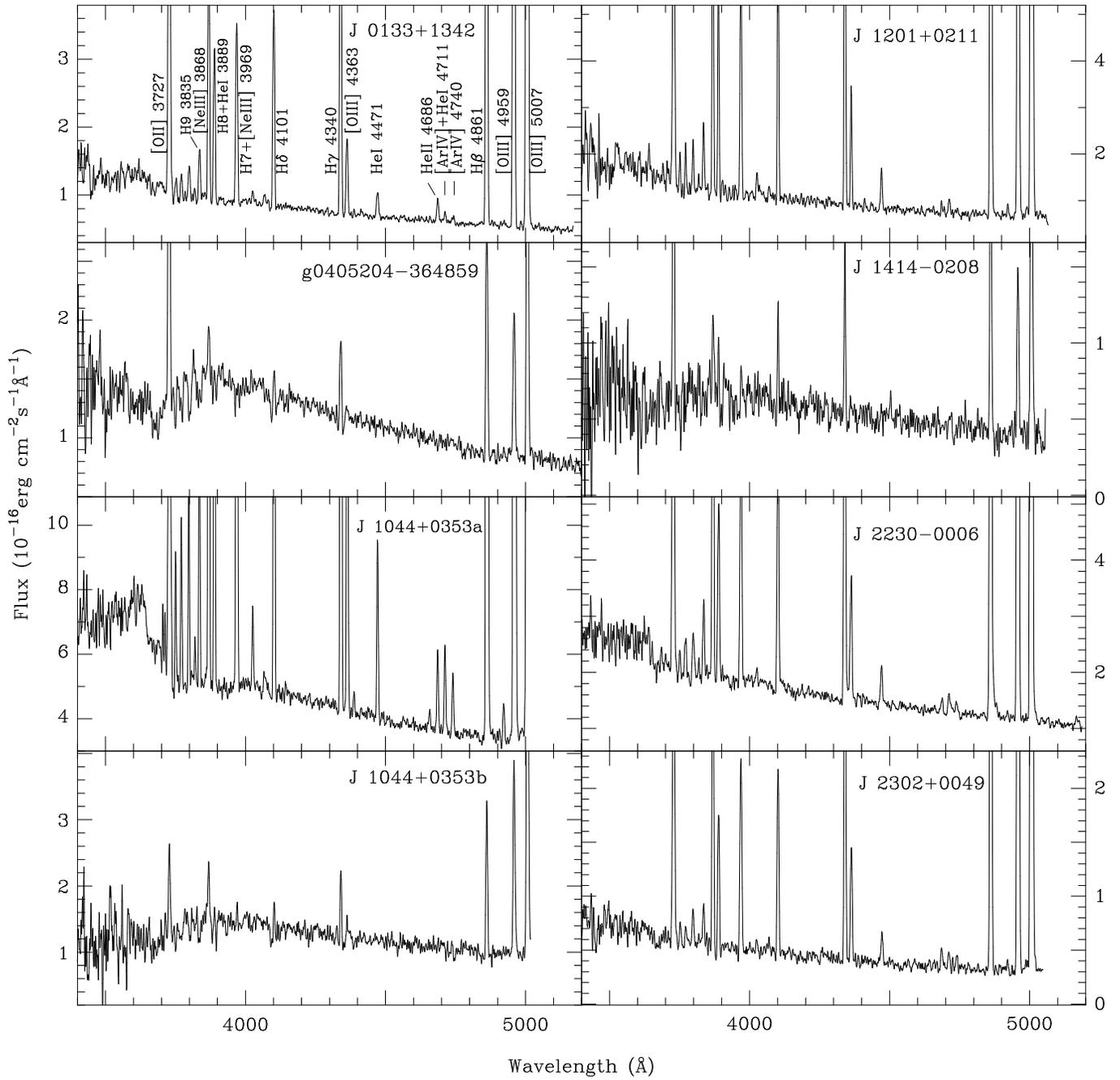}
\caption{Redshift- and extinction-corrected spectra of 8 \h2\
regions in the metal-poor galaxies studied with the ESO 3.6m telescope.  
Line identifications are shown for the first galaxy in our list.}
\label{fig_spectra}
\end{figure*}
%%%%%%%%%%%%%%%%%%%%%%%%%%%%%%%%%%%%%%%%%%%%%%%%%

The observing conditions were photometric during all nights.
For the spring observations, we used the grism $\#14$ and the grating 
600 gr/mm. The long-slit of dimensions 1\arcsec$\times$300\arcsec\ was centered 
on the brightest part of each galaxy.
The chosen instrumental setup allowed a wavelength coverage of 
$\lambda$$\lambda$3200--5083, a spectral resolution of $\sim$6.2~$\AA$ (FWHM),
and a spatial scale of 0\farcs314 pixel$^{-1}$ along the slit for the chosen
2$\times$2 pixel binning.

During the autumn observations, the grism $\#07$ and the grating 600 gr/mm were used,
resulting in a wavelength coverage of $\lambda$$\lambda$3250--5200.
These observations were carried out with a 1\farcs2$\times$300\arcsec\ slit
centered on the brightest part of each galaxy. The spectral resolution and spatial 
scale along this slit were $\sim$6.2~$\AA$ (FWHM) and 0\farcs157 pixel$^{-1}$, 
respectively.

All galaxies were observed at low airmass ($\la$ 1.2) or along
the parallactic angle, therefore no corrections for atmospheric refraction 
were applied.
The total exposure per target ranged between 40 and 60 min and was
split up in two to three subexposures to allow for a more efficient 
cosmic ray rejection. 
The seeing varied between 0\farcs8 and 1\farcs5.
Three spectrophotometric standard stars were observed during each night
for flux calibration.
The journal of the observations is given in Table~1.

In Fig.~1, we reproduce the slit position on optical CCD images.
The morphology of the galaxies is illustrated using stacked
acquisition exposures (panel b) or SDSS $gri$ images (panels a and c--g).
In Table~1, we list the adopted distance to each target.
Distances were derived after correction of the measured redshifts 
for the motion relative to the Local Group centroid and the Virgocentric flow 
\citep{K86} and assuming a Hubble constant of 75 km s$^{-1}$ Mpc$^{-1}$.

We note that the slit used is in all cases narrower
than the isophotal size of the star-forming component \P25\ 
(cf. Col. 4 in Table \ref{tab:phot}).
Given that the intensity profile of the SF component 
(Fig. \ref{SBPs}) is strongly centrally peaked with an effective radius 
r$_{\rm eff,SF} \la 1\arcsec$ (Table \ref{tab:phot}, Col. 12), our long-slit spectra 
include in all but one case (J1414-0208) most of its emission. 
Therefore, our results are not affected significantly by aperture effects, 
which, for more distant targets, can dilute the equivalent widths (EWs) of emission 
lines due to inclusion of the adjacent stellar background of the LSB host galaxy.
This is probably also the case for g0405204-364859, the r$_{\rm eff,SF}$ of which was 
estimated from stacked acquisition exposures to be $<$2\arcsec.
However, given that in several BCDs nebular emission is 
more extended than \P25\ (P02), one cannot exclude that emission line 
fluxes and EWs are slightly underestimated.

The 3.6m spectra were reduced with the IRAF\footnote{IRAF is 
the Image Reduction and Analysis Facility distributed by the 
National Optical Astronomy Observatory, which is operated by the 
Association of Universities for Research in Astronomy (AURA) under 
cooperative agreement with the National Science Foundation (NSF).}
software package. This includes bias--subtraction, 
flat--field correction, cosmic-ray removal, wavelength calibration, 
night sky background subtraction, correction for atmospheric extinction, and 
absolute flux calibration of each two--dimensional spectrum.
We extracted one-dimensional spectra from the brightest part of 
each galaxy studied.
The spectra were corrected for interstellar extinction 
using the reddening curve by \citet{W58} and transformed into the  
rest-frame based on the redshift $z$, derived from several bright emission lines. 
The redshift-corrected one-dimensional spectra of the 
8 selected \h2\ regions are shown in Fig.~\ref{fig_spectra}.

Emission line fluxes were measured using Gaussian profile fitting. 
The errors in the line fluxes were calculated from the photon
statistics of the non-flux-calibrated spectra. 
These were then propagated in the 
calculations of the elemental abundance errors.

The observed fluxes of all hydrogen Balmer lines except for
the H7 and H8 lines were used to determine the interstellar extinction and 
the underlying stellar absorption. The two excluded lines were blended with other 
strong emission lines and could not therefore be used for the determination
of interstellar reddening. 
%*************************************************************
%             Tab 2_1
%*************************************************************
\begin{figure*}
\hspace*{1.0cm}\psfig{figure=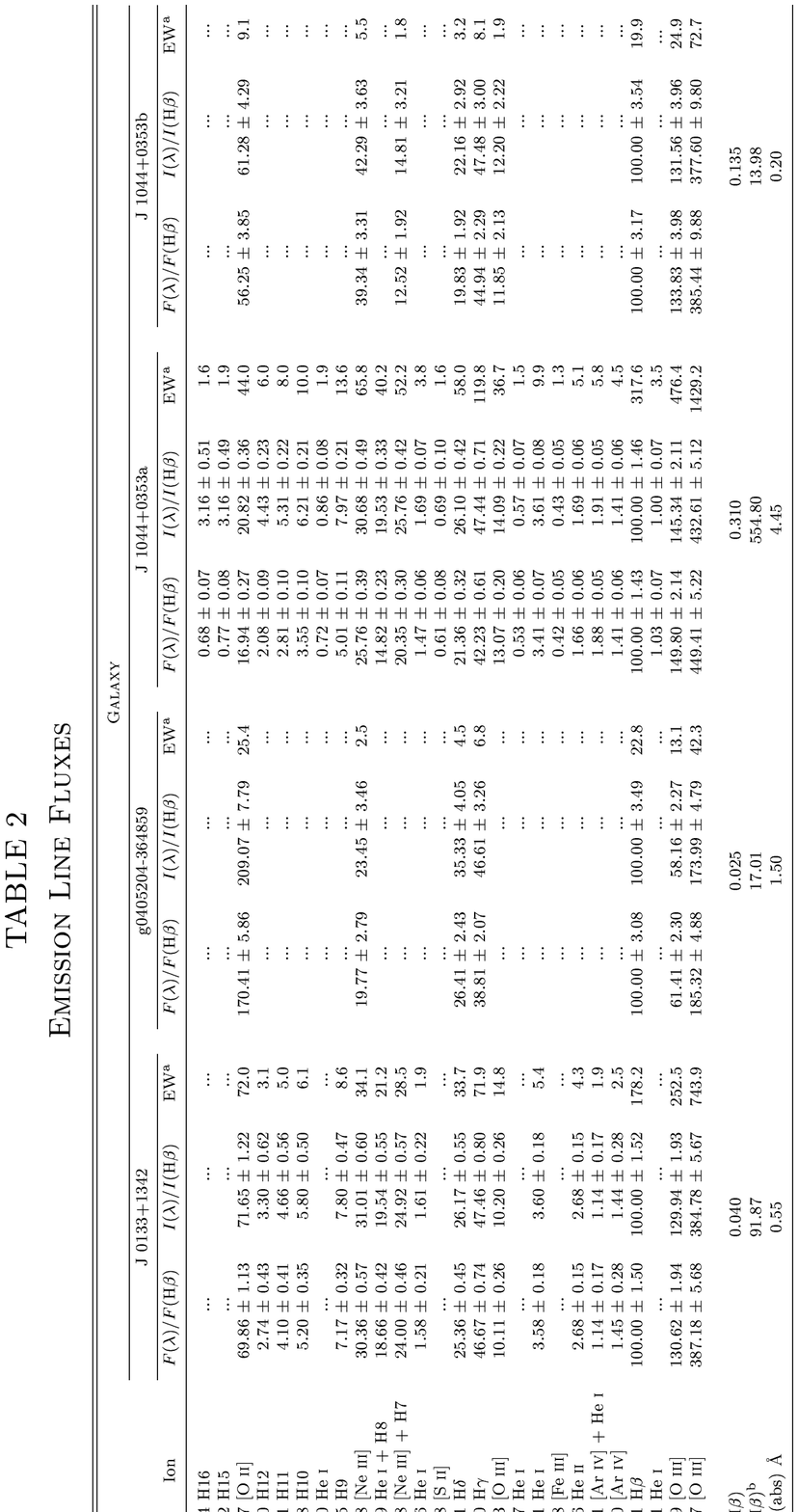,angle=-180,width=18.cm,clip=}
    \label{t2_1}
\end{figure*}

\clearpage

%*************************************************************
%             Tab 2_2
\setcounter{table}{2}
%*************************************************************
\begin{figure*}
\hspace*{1.0cm}\psfig{figure=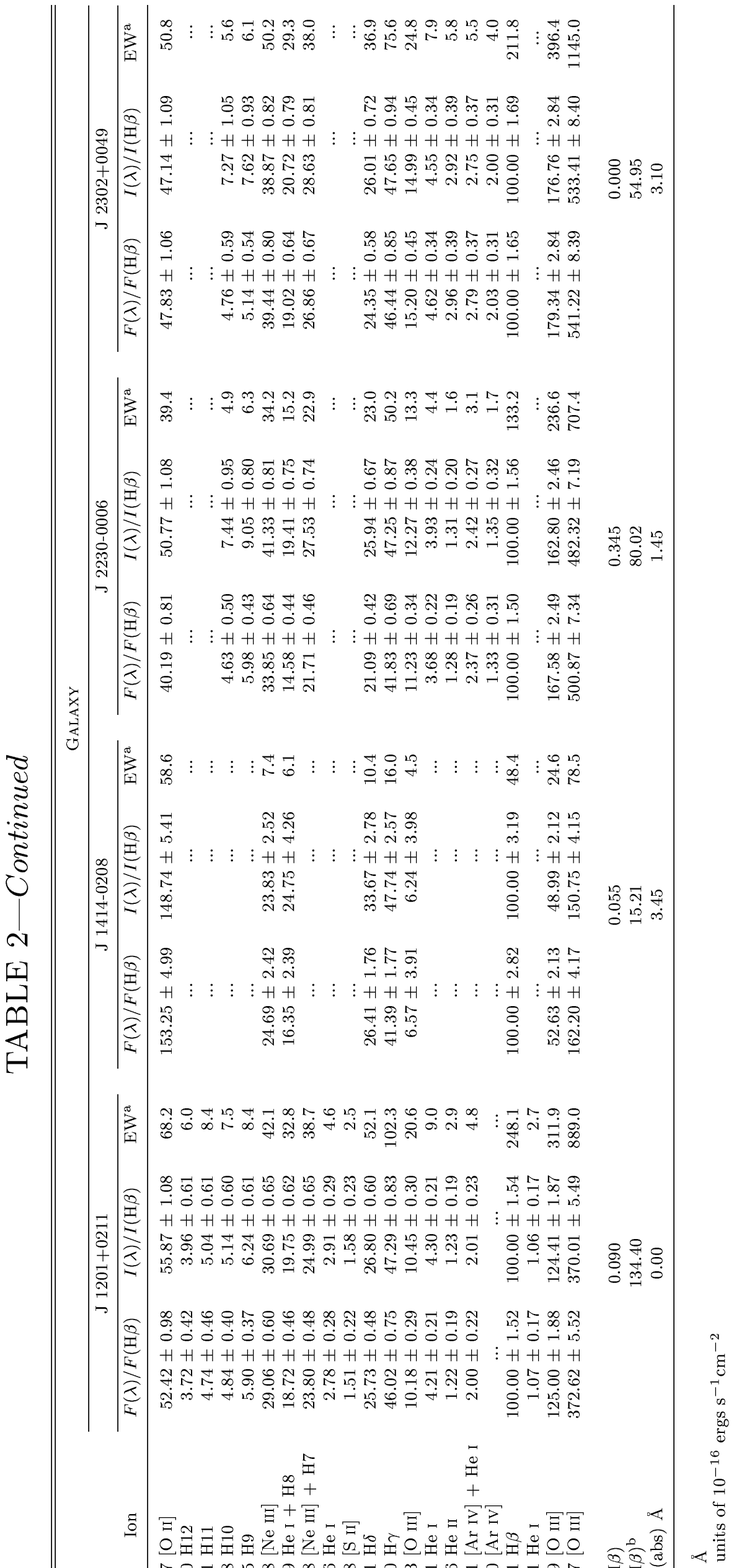,angle=-180,width=18.cm,clip=}
%\hspace*{1.0cm}\psfig{figure=Figures/tab1_2.ps,angle=-180,width=18.cm,clip=}
    \label{t2_2}
\end{figure*}

\clearpage

%*************************************************************
%             Tab 3
\setcounter{table}{3}
%*************************************************************
\begin{figure*}
\hspace*{1.0cm}\psfig{figure=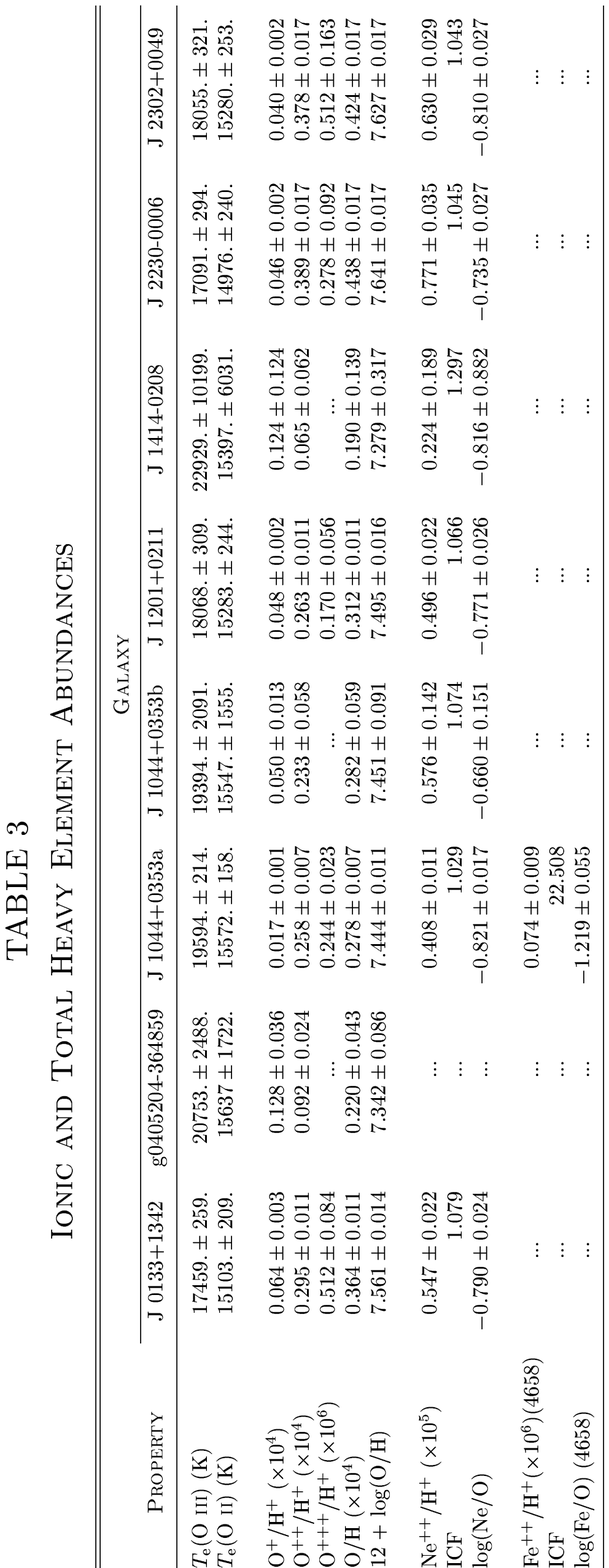,angle=-180,width=18.cm,clip=}
    \label{t3}
\end{figure*}

\clearpage

The extinction coefficient $C$(H$\beta$) and 
the equivalent width of absorption hydrogen lines EW$_{\rm abs}$ were derived by 
minimising the deviations in the corrected hydrogen emission line fluxes from the 
theoretical recombination values. In this procedure, we assumed that
EW$_{\rm abs}$ is the same for all hydrogen lines.
The observed emission line fluxes $F$($\lambda$) relative to the H$\beta$  
fluxes and those corrected for interstellar extinction and 
underlying stellar absorption $I$($\lambda$) relative to the H$\beta$, 
equivalent widths EW of emission lines,
extinction coefficients $C$(H$\beta$), 
observed H$\beta$ fluxes $F$(H$\beta$), and 
equivalent widths of the hydrogen absorption lines 
are listed in Table~2.

%%%%%%%%%%%%%%%%%%%%%%%%%%%%%%%%%%%%%%%%%%%%%%%%
%    Fig. 3 - Surface Brightness Profiles
%%%%%%%%%%%%%%%%%%%%%%%%%%%%%%%%%%%%%%%%%%%%%%%%
\begin{figure*}
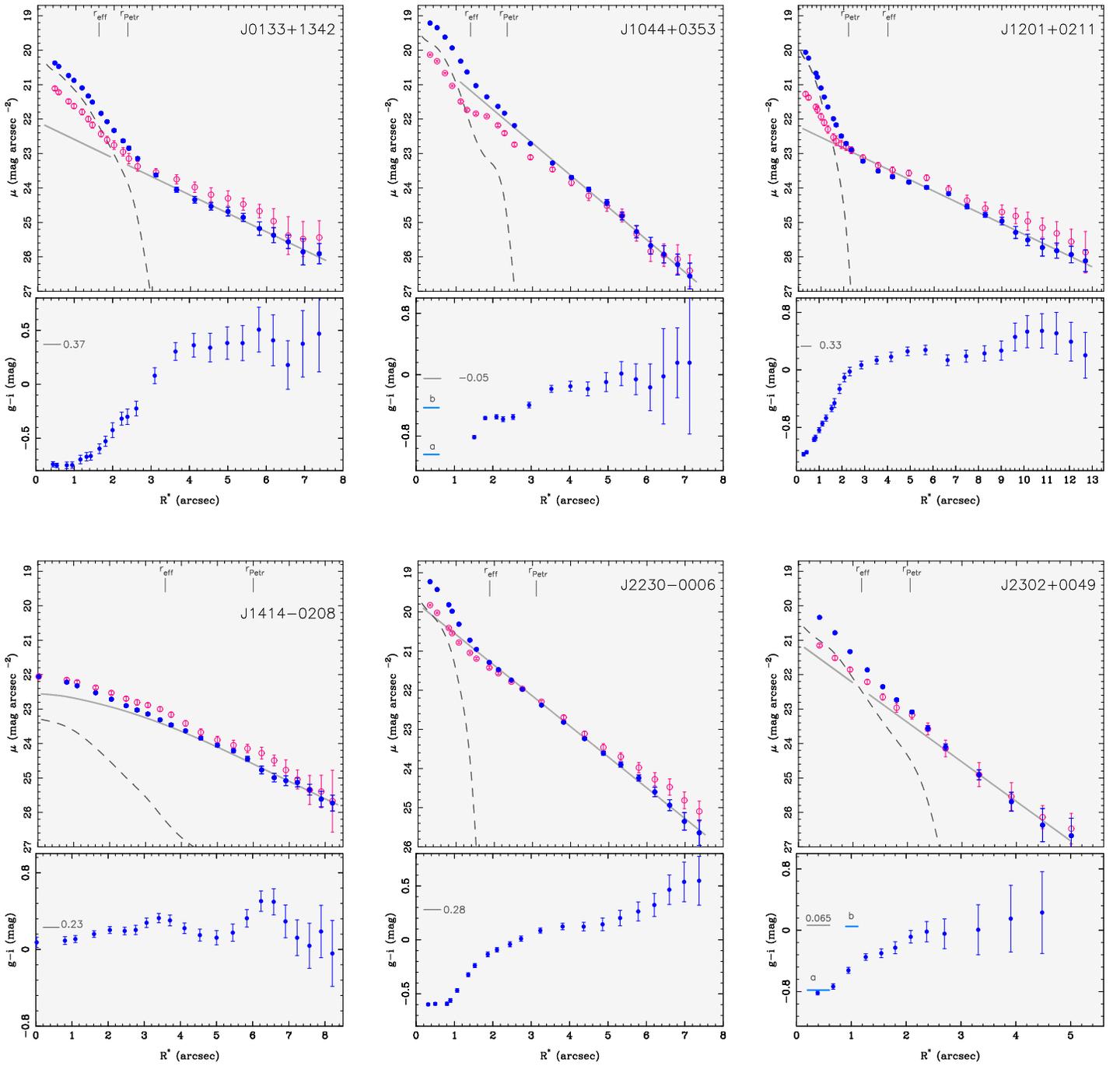

\begin{picture}(17,18)
\put(-0.5,18){{\psfig{figure=aa10028f12.ps,width=5.0cm,angle=-90.0,clip=}}}
\put(-0.5,13){{\psfig{figure=aa10028f13.ps,width=3.57cm,angle=-90.0,clip=}}}
\put(6,18){{\psfig{figure=aa10028f14.ps,width=5cm,angle=-90.0,clip=}}}
\put(6,13){{\psfig{figure=aa10028f15.ps,width=3.57cm,angle=-90.0,clip=}}}
\put(12.5,18){{\psfig{figure=aa10028f16.ps,width=5cm,angle=-90.0,clip=}}}
\put(12.5,13){{\psfig{figure=aa10028f17.ps,width=3.57cm,angle=-90.0,clip=}}}
\put(-0.5,8.5){{\psfig{figure=aa10028f18.ps,width=5.0cm,angle=-90.0,clip=}}}
\put(-0.5,3.5){{\psfig{figure=aa10028f19.ps,width=3.57cm,angle=-90.0,clip=}}}
\put(6,8.5){{\psfig{figure=aa10028f20.ps,width=5.0cm,angle=-90.0,clip=}}}
\put(6,3.5){{\psfig{figure=aa10028f21.ps,width=3.57cm,angle=-90.0,clip=}}}
\put(12.5,8.5){{\psfig{figure=aa10028f22.ps,width=5.0cm,angle=-90.0,clip=}}}
\put(12.5,3.5){{\psfig{figure=aa10028f23.ps,width=3.57cm,angle=-90.0,clip=}}}
\end{picture}
\caption{Surface brightness profiles (SBPs) of the SDSS sample galaxies (upper
  panels) in $g$ (full circles) and $i$ (open circles). 
  Solid lines show fits to the LSB host galaxy (Table \ref{tab:phot})
  according to Eq. \ref{eq:p96a} and Eq. \ref{eq:exp} for \object{J1414-0208} and
  all other objects, respectively. The surface brightness distribution of
  the SF component in the $g$ band is shown by dashed curves.
The effective r$_{\rm eff}$ and Petrosian r$_{\rm Petr}$ radii of each galaxy 
are indicated by vertical lines.
  $g-i$ colour profiles (lower panels) have been computed by subtraction of
  the $i$ SBPs from the $g$ SBPs. The mean $g-i$ colour of the LSB host galaxy within
  the fitting range (Col. 10 in Table \ref{tab:phot}) is indicated by the
  horizontal lines at the left part of each diagram. 
  The mean colours of the regions {\sf a} and {\sf b} 
  in \object{J1044+0353} and \object{J2302+0049} are depicted by
  horizontal lines on the left-hand side of the corresponding diagrams.
}
\label{SBPs}
\end{figure*}

\begin{table*}
\caption{\label{decomp_res}Structural properties of the SDSS sample galaxies
  in the $g$ band.
}
\label{tab:phot}
\begin{tabular}{lccccccccccc}
\hline
\hline
Name  & $\mu_{\rm E,0}$ & $M_{\rm SF}$ & $R_{\rm SF}$     & $m$ & r$_{\rm eff}$   & r$_{\rm Petr}$
& $CI$ & $R_{\rm host}$/r$_{\rm eff}$ & fit range & t(SFH1,2) &
r$_{\rm eff,SF}$\\
       & $\alpha$      & $M_{\rm host}$ & $R_{\rm host}$ & $M$ & $\mu_{\rm eff}$ & m$_{\rm Petr}$ &
$\eta$ &  $l_{\rm SF}$($R_{\rm host}$,total) & ($g-i$)$_{\rm host}$ & M$_{\star}$(SFH1,2)
&  $m$-$M$ \\
(1)  &  (2)  & ( 3) &  (4) & (5) & (6) & (7) & (8) & (9) & (10) & (11) &
(12) \upperspace\lowspace \\
\hline
% ++++++++++++++++++++++++++++++++++++++++++++++++++++++++++++++++++++++++++++++++++++++++
J 0133$+$1342 & 22.1$\pm$0.1  & --14.2 & 0.47 &   17.9 & 0.30  & 0.43  & 0.74 &
3.13 & 3\farcs9 -- 7\farcs2 & 1.0,2.7 & 1\farcs1\\
     & 0.37$\pm$0.02       & --14.3 & 0.93 & --15.0 & 20.98 & 18.4  & 2.25 & 
0.47,0.53 & 0.37$\pm$0.09 & 1.9,3.0 & 32.887\\
\hline
J 1044+0353 & 19.9$\pm$0.2  & --15.0 & 0.60 & 17.2 & 0.35  & 0.57 & 0.78 &
 3.63 & 3\farcs2--7\farcs5 & 0.1,1.0 & 0\farcs8\\
      & 0.29$\pm$0.01  &  --15.9   & 1.27 & --16.3  & 19.9 & 17.5 & 1.4 & 
0.69,0.70 & --0.05$\pm$0.13 & 0.4,2.2 & 33.538\\
\hline
J 1201+0211 & 22.2$\pm$0.1  & --11.7 & 0.14 & 17.4   & 0.27  & 0.15 & 0.92 & 
2.00 & 4\farcs6--13\farcs1 & 1.0,2.5 & 0\farcs8\\
     & 0.23$\pm$0.01    & --12.7 & 0.54 & --13.3 & 22.4  & 18.5 & 5.2 & 
0.72,0.79 & 0.33$\pm$0.14 & 1.9,2.5 & 30.731\\
\hline
J 1414-0208 & 21.6$\pm$0.1  & --10.8 & 0.24 & 18.1 & 0.38  & 0.67 & 0.87 & 
1.80 & 3\arcsec--8\farcs4 & 0.6,2.0 & 1\farcs8\\
     & 0.24$\pm$0.01    & --13.4 & 0.69 & --13.7 & 22.8 & 18.3  & 0.89 & 
0.92,0.94 & 0.23$\pm$0.1 & 0.4,0.7 & 31.809\\
\hline
J 2230-0006 & 19.8$\pm$0.1  & --13.0 & 0.19 & 17.0   & 0.23  & 0.37  & 0.94 & 
3.36 & 3\arcsec--7\farcs7 & 0.8,2.2 & 0\farcs7\\
     & 0.17$\pm$0.01     & --14.8 & 0.75 & --15.0 & 20.33 & 17.31 & 1.45 &
0.84,0.85 & 0.28$\pm$0.15 & 2.3,2.9 & 31.981\\
\hline
J 2302+0049 & 20.8$\pm$0.3  & --15.9  & 1.33 & 18.6   & 0.73  & 1.33  & 0.58 & 
2.85 &  2\farcs4--5\farcs2  &  0.3,1.4 & 0\farcs8\\
     & 0.58$\pm$0.03      & --16.4  & 2.09 & --17.0 & 20.86 & 18.87 & 1.37 &  
0.63,0.65 & 0.07$\pm$0.2    & 1.8,4.8 & 35.645\\
\hline
% ++++++++++++++++++++++++++++++++++++++++++++++++++++++++++++++++++++++++++++++++++++++++
\hline
\end{tabular}
\end{table*}
\setcounter{footnote}{0}

%%%%%%%%%%%%%%%%%%%%%%%%%%%%%%%%%%%%%%%%%%%%%%%%
%    Fig. 4 - Colour maps
%%%%%%%%%%%%%%%%%%%%%%%%%%%%%%%%%%%%%%%%%%%%%%%%
\begin{figure*}[t]
\begin{picture}(17,12)
\put(0.5,0){{\psfig{figure=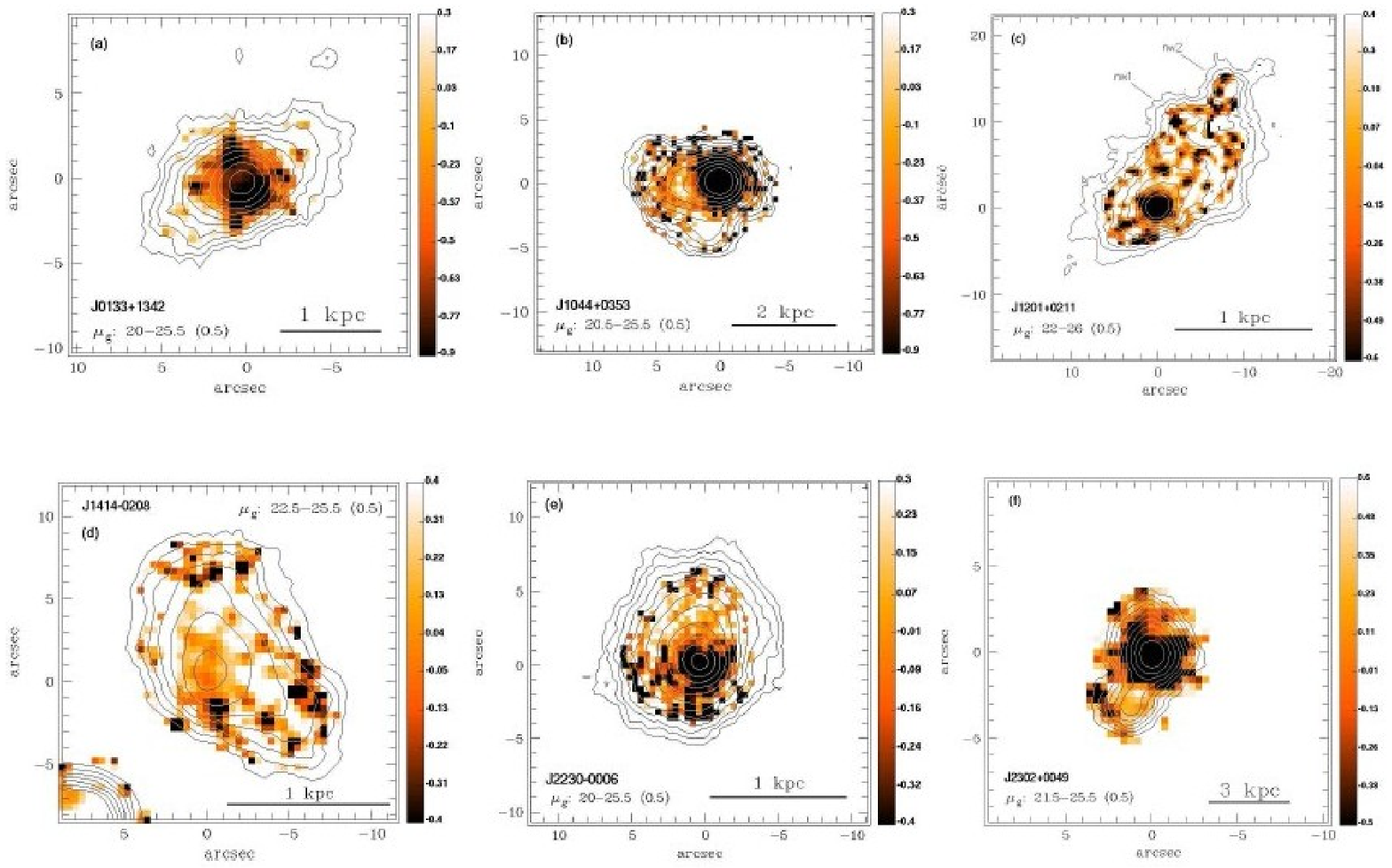,width=17cm,angle=0.0,clip=}}}
\end{picture}
\caption{$g$-$i$ colour maps of the SDSS sample galaxies displayed in the colour
  interval indicated by the vertical bars to the right of each panel. 
   The surface brightness range and the spacing of the superposed $g$ band
  contours are indicated. 
North is to the top and east to the left.}
\label{colour_maps}
\end{figure*}

% ===========================================================
\section{Results \label{results}}
\subsection{Chemical abundances \label{spectra}}
% ===========================================================

The electron temperature $T_{\rm e}$ and the ionic and total heavy element abundances
were derived following \cite{Izotov2006-SDSS}. In particular, for the O$^{2+}$
and Ne$^{2+}$ ions, we adopted the temperature $T_{\rm e}$(O {\sc iii})
derived from the [O {\sc iii}] 
$\lambda$4363/($\lambda$4959 + $\lambda$5007) emission line ratio. 
The O$^+$ and Fe$^{++}$ abundances were derived with the temperature 
$T_{\rm e}$(O {\sc ii}), which was obtained from the relation between 
$T_{\rm e}$(O {\sc iii}) and
$T_{\rm e}$(O {\sc ii}) of \citet{Izotov2006-SDSS}.

The 3.6m telescope spectra covered the blue wavelength region,
thus the [S {\sc ii}]$\lambda$6717,6731 emission lines, usually 
used for the determination of the electron number density, were 
not observed. 
Therefore for abundance determinations, we adopted  
$N_{\rm e}$ = 100 cm$^{-3}$. 
The derived abundances are insensitive to the precise value of the electron
number density in the low-density limit ($N_{\rm e}$ $\la$ 10$^3$ cm$^{-3}$),
which is applicable for the H {\sc ii} regions considered here.

The electron temperatures $T_{\rm e}$(O {\sc iii}) and $T_{\rm e}$(O {\sc ii}) 
for the high and low-ionization zones in H {\sc ii} regions, respectively,
the ionization correction factors ($ICF$s), and the 
ionic and total heavy element abundances for
oxygen, and, where possible, neon and iron are listed in Table 3.

The derived oxygen abundance \oh\ of the sample galaxies is in the range
between 7.3 and $\sim$7.6, the objects investigated here therefore 
belong to the rare class of extremely metal-deficient emission-line galaxies with 
an oxygen abundance $\la$7.6 \citep{TI05}. 
The N/O and Fe/O abundance ratios are consistent within the errors with the ratios
obtained for other metal-deficient galaxies \citep{IT99}.

Comparison of 3.6 m and SDSS data for the common three \h2\ regions shows that
the 3.6m oxygen abundances are lower by $\sim$0.04 dex than those in the
papers of \citet{K03,K04b}.
These relatively small discrepancies can be explained by 
differences in the data acquisition (e.g. telescope, sky transparency) and
methods employed in the chemical abundance determination.

It is worth pointing out that all known XBCDs with oxygen abundance
determinations in multiple positions over their optical extent 
show little, if any, abundance variations.
Therefore, the oxygen abundance derived in one \h2\ region of these systems
may be considered to within $\sim$0.1 dex as representative 
of the ionized interstellar medium (ISM) over the entire galaxy.
Examples are SBS 0335-052E, which differs by less than 0.1 dex in the 
oxygen abundance of its central SF clusters \#1,2 and its peripheral 
SF region \#7 \cite[cf.][]{Izotov1997-SBS0335E,Papaderos2006-SBS0335}, 
and I Zw 18 of essentially equal oxygen abundance in both its 
northwestern and southeastern SF region \citep{Izotov1999_He_IZw18_SBS335}.
Several other examples may be found in e.g. \cite{IzotovThuan2007-MMT}.

% =============================================
\subsection{Photometric analysis \label{photo}}
% =============================================
\subsubsection{Derivation and general properties of surface brightness
  profiles \label{photo1}}
Surface brightness profiles (SBPs) for the SDSS galaxies were derived 
using method {\tt iv} introduced in P02.
This SBP determination algorithm was developed especially for the study of 
irregular SF galaxies, presenting strong intensity and colour gradients,
and has been successfully applied both to nearby BCDs 
\citep[P02,][]{Noeske2003-NIR} and 
compact SF galaxies at intermediate redshift \citep{Noeske2006-UDF}.

Surface photometry was carried out in the $g$ and $i$ bands. As is evident from
Fig. ~\ref{SBPs}, SBPs reach in all cases a surface brightness of 
$\sim$26 $g$ \sbb, i.e. they allow the detection and quantitative study of 
the faint and more extended LSB host galaxy underlying the SF component.
It is well established from earlier work (P96b) that
the host dominates the line-of-sight intensity for $\mu \ga$ 24.5 $B$ \sbb. 
This is also true for dwarf irregulars (dIs), which, typically, 
do not show an appreciable SF activity in their LSB periphery 
\citep{PT1996,MK1998,YH1999,vanZee2000}.

The SBPs of most sample galaxies reveal a two-component structure 
that is typical of BCDs i.e. an exponential slope 
at large radii which characterises the LSB emission, and 
a central luminosity excess, which can be attributed to the SF component.
These two distinct stellar populations are also reflected in colour maps (Fig. \ref{colour_maps}) 
and colour profiles (Fig. ~\ref{SBPs}), which display a considerable colour contrast
between the SF center and the LSB periphery (cf. \object{J0133+1342,
  J1201+0211, J2230-0006}), as typically found for BCDs 
\citep[P96b, P02,][]{Cairos2001b,GildePazMadore2005}.
The only galaxy in our sample undergoing weak SF activity with 
a comparatively low EW(\hb) ($\sim$50 $\AA$) and an almost  
exponential SBP all the way to its center is \object{J1414-0208}. 
Not surprisingly, this system also exhibits negligible colour gradients 
and subtle colour variations of an amplitude less than 0.4 mag.

SBPs were decomposed into separate profiles representing 
the luminosity of the SF and the LSB components 
following the procedure described in P02.
These two components are shown in each SBP (Fig. \ref{SBPs}) as solid and
dashed lines, respectively.
We chose not to apply the 3-component scheme used in P96a,
which consists of an exponential, S\'ersic, and Gaussian component, 
which represent the host galaxy and the SF component at both intermediate
and small radii, respectively.

This was mainly because the central Gaussian component does not 
correlate with other galactic structural properties and 
is always far narrower than the isophotal radius of the SF component, termed 
by P96a as \emph{plateau radius} \P25.
Instead, we consider only the total emission from the SF
component, i.e. the emission in excess of the exponential fit to the LSB host.

In all but one of the SDSS galaxies studied, the LSB host 
was approximated by an exponential fitting function of the form:
\begin{equation}
I(R^*) \equiv I_{\rm exp} = I_0\,\exp\left( -\frac{R^*}{\alpha}\right), 
\label{eq:exp} 
\end{equation}
where $I_0$ and $\alpha$ denote the central intensity and the exponential
scale length of the host, and \rr\ is the photometric radius.

In the case of \object{J1414-0208}, Eq. \ref{eq:exp} is unsuitable because
at intermediate radii it predicts a higher intensity than observed. 
Evidently, the outer exponential profile of the LSB component cannot
continue all the way to \rr=0\arcsec\ but probably flattens at small radii.
Therefore, following P96a, we fitted the LSB host of \object{J1414-0208}
with a modified exponential model involving a mild central flattening of the form:
\begin{equation}
I(R^*) = I_{\rm exp} \cdot
\big[1-\epsilon_1\,\exp(-P_3(R^*))\big],
\label{eq:p96a} 
\end{equation}
where $P_3(R^*)$ is defined as
\begin{equation}
P_3(R^*) =
\left(\frac{R^*}{\epsilon_2\,\alpha}\right)^3+\left(\frac{R^*}{\alpha}\,\frac{1-\epsilon_1}{\epsilon_1}\right).
\label{eq:p96b} 
\end{equation}

For radii \rr$\la \epsilon_2 \cdot \alpha$, 
the distribution given by Eq. \ref{eq:p96a} 
depends on the central depression $\epsilon_1=\Delta I/I_0$ relative to the pure
exponential profile $I_{\rm exp}$. 
Given that $\epsilon_2$ is typically $\approx 3\cdot \epsilon_1$ 
\citep[cf.,][]{Fricke2001-T1214,Noeske2003-NIR,Noeske2005-NIR}, 
Eq. \ref{eq:p96a} can be simplified to a 3-parameter formula. 
In the case of \object{J1414-0208}, 
we adopt a moderate central flattening $\epsilon_1=0.6$, which corresponds 
to a central surface brightness and total magnitude that are 0.55 mag and 0.2
mag fainter, respectively, than those predicted by fitting Eq. \ref{eq:exp}.

% =========================================================================================
\subsubsection{Spatial extent of the star-forming component and the LSB host\label{photo2}}
% =========================================================================================
One of our principle aims is to use surface photometry to determine the
structural properties of BCDs/XBCDs and study the influence of the gravitational 
potential of their LSB host on their star formation process (see also P96b).
For this purpose, it is important to first disentangle the LSB and the SF component
and determine their respective mass fractions and spatial extents.
To this end, P96a introduced the isophotal radii \E25\ and \P25\ of these two components 
at an extinction-corrected surface brightness level of 25 $B$ \sbb. 

A practical advantage of \P25\ is that it can be easily derived from a 
2-component SBP decomposition and shows little dependence on the 
luminosity of the SF component, which in BCDs may strongly vary on 
a timescale of a few $10^7$ yr.

This is not the case for either the effective radius or concentration
indices evaluated from the radii enclosing different fractions of the total flux
(for example, the ratio $\log(R_{80}/R_{20})$ of the radii containing
80\% and 20\% of the total light). These quantities depend primarily 
on the luminosity rather than on the spatial extent of the SF component
\citep[see e.g. P96a and][P06]{BartonvanZee2001}.
%
%%%%%%%%%%%%%%%%%%%%%%%%%%%%%%%%%%%%%%%%%%%%%%%%
%    Fig. 5 (g2b1.okm)
%%%%%%%%%%%%%%%%%%%%%%%%%%%%%%%%%%%%%%%%%%%%%%%%
\begin{figure}
\hspace*{-0.0cm}\psfig{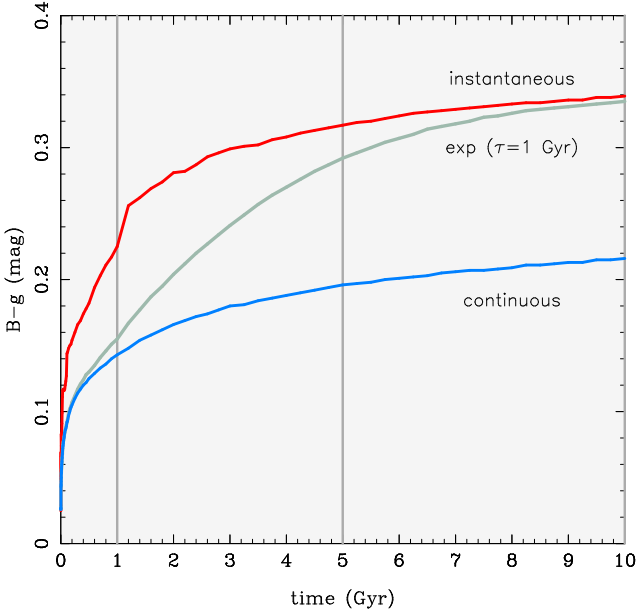}
\caption{Evolution of the $B-g$ colour index as a function of time for three star
  formation histories: {\lvss SFH1} (instantaneous), {\lvss SFH2} (exponentially decreasing 
star formation rate with an e-folding time of 1 Gyr) and {\lvss SFH3}
  (continuous, constant star formation rate). 
}
\label{B-g}
\end{figure}
%%%%%%%%%%%%%%%%%%%%%%%%%%%%%%%%%%%%%%%%%%%%%%%%%

To be able to compare directly the photometric properties, 
e.g., \P25, \E25 and magnitudes, of the present galaxy
sample in the SDSS $g$ with the available literature data in $B$ 
it is necessary to first infer the $g$ surface brightness that corresponds 
to a $B$ surface brightness of 25 \sbb.
For this, we study the time evolution of the $B-g$ index for three 
different star formation histories (SFHs):
an instantaneous SF episode ({\lvss SFH1}), an exponentially decreasing star
formation rate with an e-folding timescale of 1 Gyr ({\lvss SFH2}), and a 
constant, continuous SF ({\lvss SFH3}). 
Calculations were carried out with the PEGASE 2.0 code \citep{FRV1997}
and refer to a stellar population with a fixed stellar metallicity
of $Z$=0.001 and a Salpeter initial mass function with lower and upper mass 
limits of 0.1 and 100 \msun, respectively. 

Figure \ref{B-g} shows that the $B-g$ index for {\lvss SFH1} and {\lvss SFH2}
differs by less than 0.1 mag after $t\sim$1 Gyr, and that both SFHs imply 
a $B-g \approx $ 0.3 mag for t$\geq$5 Gyr. 
Since these two models are probably the most appropriate 
approximations of the SFH of the SF component
and that of the LSB host, we assume in the following the relation $m_B
\approx m_g + 0.3$. 

% ==========================================================================
\subsubsection{Photometric properties of the sample galaxies \label{photo3}}
% ==========================================================================
Table \ref{tab:phot} lists photometric quantities for the six SDSS galaxies in
our sample. All tabulated values are corrected for Galactic absorption 
based on the extinction maps by \cite{Schlegel1998} available in the
NED\footnote{NASA/IPAC Extragalactic Database;\\ {\tt http://nedwww.ipac.caltech.edu}}.
Column 2 of Table \ref{tab:phot} gives the \emph{extrapolated} central surface
brightness $\mu_{\rm E,0}$ (\sbb) 
and the exponential scale length $\alpha$ (kpc) of the LSB host galaxy, obtained
by fitting Eq. \ref{eq:exp} to the outer part of each SBP.  
In Col. 3, we tabulate the absolute $g$ band magnitude of the SF
component $M_{\rm SF}$ and that of the LSB host $M_{\rm host}$ within 
their respective isophotal radii \P25\ and \E25\ (Col. 4) at a surface
brightness level of 24.7 $g$ \sbb\ (see discussion in Sect. \ref{photo2}). 
In Col. 5, we tabulate the apparent and absolute magnitude
of each galaxy as derived from SBP integration.
Column 6 lists the effective radius r$_{\rm eff}$ (kpc) and 
the mean surface brightness $\mu_{\rm eff}$ inside r$_{\rm eff}$,
and in Col. 7 we list the Petrosian radius r$_{\rm Petr}$ (kpc), which is 
defined to be the radius at which the Petrosian $\eta$ function 
\citep{Petrosian1976} decreases 
to a value of 1/3 \citep[see e.g.][]{Takamiya1999}, and the apparent
magnitude m$_{\rm Petr}$ within r$_{\rm Petr}$.

The concentration index $CI$, defined by P96a to be 
1-(\P25/\E25)$^2$ is tabulated in Col. 8. 
This quantity measures the fractional area of the SF component 
and attains a maximal value for systems containing a single, compact 
SF region. In the same column, we list the S\'ersic exponent
$\eta$ obtained from fitting a S\'ersic model of the form 
$I_0 \exp(-{\rm R^{\star}}/\alpha)^{1/\eta}$ to each $g$ SBP.
Column 9 lists the ratio \E25/r$_{\rm eff}$ and the 
luminosity contribution $l_{\rm SF}$ of the host galaxy 
both within \E25\ and with regard to the total galaxy emission. 
In Col. 10, we tabulate the fitted radius range of the LSB host 
and the mean $g-i$ colour within that radius interval.
The age of the host galaxy in Gyr, estimated from its colour 
using PEGASE 2.0 models for {\lvss SFH1} and {\lvss SFH2}, and 
its corresponding present stellar mass in $10^7$ \msun\ are given in Col. 11.
Column 12 lists the effective radius r$_{\rm eff,SF}$ of the SF component 
in arcsec and the distance modulus $m-M$ of each target.

%%%%%%%%%%%%%%%%%%%%%%%%%%%%%%%%%%%%%%%%%%%%%%%%
%    Fig. 6 (evolution of the g-i colour)
%%%%%%%%%%%%%%%%%%%%%%%%%%%%%%%%%%%%%%%%%%%%%%%%
\begin{figure} 
\hspace*{-0.0cm}\psfig{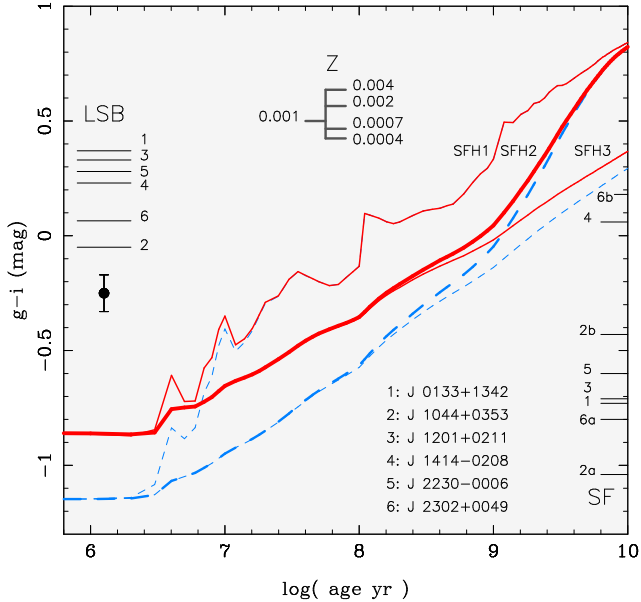}
\caption{Evolution of the $g-i$ colour index as a function of time for the star
  formation scenarios {\lvss SFH1} through {\lvss SFH3} (Sect. \ref{photo2})  
based on Pegase 2.0 models involving purely stellar emission (solid lines) as well 
as stellar and ionized gas emission (dashed lines).
The average colour shift within 1$\leq$t (Gyr)$\leq$3 
that a variation in the stellar metallicity between
0.0004 and 0.004 would produce is indicated.
Horizontal lines on the left-hand side of the diagram indicate the mean 
$g-i$ colour of the LSB host within the fitted radius range 
(Table \ref{tab:phot}, Col. 10) along with a typical 1$\sigma$ uncertainty.
Likewise, the measured $g-i$ colour in the SF component is depicted by horizontal lines
at the r.h.s. part of the diagram.}
\label{g-i-evol}
\end{figure}
%%%%%%%%%%%%%%%%%%%%%%%%%%%%%%%%%%%%%%%%%%%%%%%%%

\begin{figure} 
\hspace*{-0.0cm}\psfig{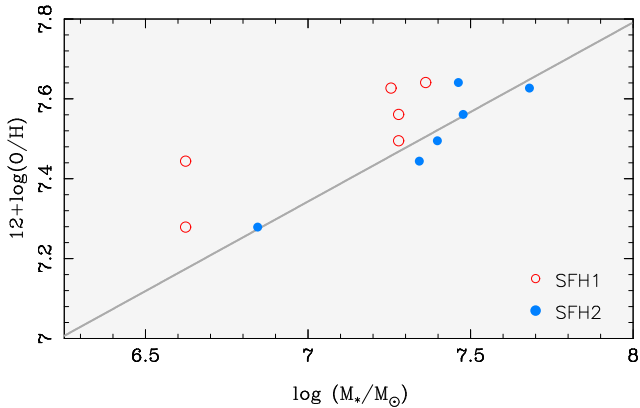}
\caption{Relation between the oxygen abundance \oh\ and the 
stellar mass M$_{\star}$ of the LSB host of our SDSS sample galaxies. 
Stellar masses were derived using Pegase 2.0 models on the base of the star formation
histories {\lvss SFH1} (open symbols) and {\lvss SFH2} (filled symbols).
The diagonal line shows a linear fit to the oxygen abundance for stellar 
masses estimated assuming {\lvss SFH2}.
}
\label{mz}
\end{figure}

In Fig. \ref{mz}, we plot the stellar mass M$_{\star}$ of the LSB host of 
our sample galaxies, derived on the basis of their star formation histories 
{\lvss SFH1} and {\lvss SFH2} and oxygen abundance. 
A clear trend (Spearman's rank correlation coefficient: 0.83) 
of the form
\begin{equation}
{\rm 12+log(O/H)} = (4.2\pm 0.6) + (0.45\pm 0.09) \log_{10} {\rm M}_{\star}
\end{equation}
between these two quantities is evident for masses derived
adopting {\lvss SFH2}. The derived slope between gas-phase metallicity and 
stellar mass, $Z\propto M_{\star}^{0.45\pm0.09}$, is in good agreement with the 
relation $Z\propto M_{\star}^{0.4}$ between stellar metallicity and mass 
predicted by \cite{DekelWoo2003} and slightly steeper than the empirical 
relation $Z\propto M_{\star}^{0.3}$ inferred by \cite{Lee2006} from the 
integrated 4.5$\mu$m luminosity of nearby dwarf irregular galaxies. 

%%%%%%%%%%%%%%%%%%%%%%%%%%%%%%%%%%%%%%%%%%%%%%%%
%    M*-Z relation
%%%%%%%%%%%%%%%%%%%%%%%%%%%%%%%%%%%%%%%%%%%%%%%%

% ======================================================
\section{Remarks on individual galaxies \label{objects}}
% ======================================================

% ==================
\object{J0133+1342}: 
% ==================
This system contains an unresolved high-surface brightness 
SF region in the central part of a moderately irregular LSB host.
The absolute $g$ magnitude of the host, $M_{\rm g} \approx -14.3$ mag, 
translating to a $B$ band magnitude $M_{\rm B} \approx -14$ mag, 
places \object{J0133+1342} well within the range of dwarf galaxies. 
The central surface brightness ($\mu_0 \approx 22$ \sbb) and 
exponential scale length ($\alpha =0.37$ kpc) of the host are 
typical of BCDs. This is also the case for the isophotal radius
(\P25$\approx$0.5 kpc) and the fractional luminosity of the SF component ($\sim$0.5).
The blue $g-i$ colour (--0.73 mag) within the central 3\arcsec$\times$3\arcsec\
of the SF component indicates a substantial young stellar population.
This is also suggested by the strong nebular emission of \object{J0133+1342} 
with an EW of 178 $\AA$ and 743 $\AA$ for the H$\beta$ and 
[O {\sc iii}] $\lambda 5007$ lines, respectively.
In agreement with the general trend observed for BCDs, this galaxy shows a gradual 
colour increase with increasing galactocentric radius and an almost constant,
reddish colour in its LSB host. 

In Fig. \ref{g-i-evol}, we compare the colours of \object{J0133+1342} with
the expected $g-i$ colour evolution for the star formation scenarios {\lvss SFH1}
through to {\lvss SFH3}.
In addition to models involving purely stellar emission (solid curves), 
we show models including ionized gas emission (dashed curves). 
The observed colours in the LSB host and SF component of the SDSS galaxies
studied are depicted by horizontal lines in the left-hand side and 
right-hand side of the figure, respectively.
From Fig. \ref{g-i-evol}, it can be seen that the mean LSB colour of 
0.37$\pm$0.08 mag for \rr$\geq$0.7 kpc translates into 
an age of between 1 and 2.5 Gyr for \sfha\ and
\sfhb, respectively, suggesting a moderately evolved stellar population.
The blue colours of the SF component correspond to a burst age of between 3 and 7 Myr
(\sfha\ assumed) with values in the lower range preferable because of the non-detection
of Wolf-Rayet spectral features for this system.

The oxygen abundance that we derive from the 3.6m ESO telescope spectra,
\oh=7.56$\pm$0.01, compares well with the value of 7.60$\pm$0.03 inferred 
by \citet{K03} from SDSS spectra, and with the value of 
\oh=7.55$\pm$0.04 that we derive from the same data following the
prescriptions of \citet{Izotov2006-SDSS}.

\smallskip
\object{g0405204-364859}: This nearby system ($D\approx 11$ Mpc), identified in 
the 6dFGRS survey, has never been studied previously. 
Because of its southern declination, it was not included in the SDSS survey, 
and archival data also do not exist; no quantitative statements can therefore
be made about its photometric structure.
On the coadded acquisition exposure (90 sec in $V$; Fig.~1), 
it appears to be a compact source with an overall regular morphology and 
elliptical LSB host with a possible faint extention in its southwestern direction. 

The spectrum of this system is of low signal-to-noise ratio, yet 
clearly reveals weak emission lines superimposed on a blue stellar continuum. 
Its oxygen abundance, determined to be 7.34$\pm$0.09, makes it one of the 
most metal-poor XBCDs discovered to date.

\smallskip
% =================================================
\object{J1044+0353}: 
% =================================================
This object is likely a pair of physically associated SF regions 
with velocity and projected linear separation of $\sim$30 \kmsec and 0.8 kpc, respectively.
The brighter, eastern SF region {\lvss a} displays extremely blue colours 
($g-i\la -1$ mag within a 3\arcsec$\times$3\arcsec\ region), which are only 
reproducible by a photoionized region model involving strong nebular emission,
in addition to a very young stellar population.
Indeed, our spectra reveal nebular lines with an EW of 318 $\AA$ and 1420 $\AA$
for the \hb\ and \o5007, respectively, implying an intense, ongoing SF episode. 
Our photometry indicates that more than 80\% of the total emission in excess of
the LSB host of \object{J1044+0353} originates in region {\lvss a}.

In contrast, the faint eastern region {\lvss b}, reflected on a weak
bump in the SBPs at \rr$\approx$2\farcs5, exhibits weak nebular emission 
with an EW(\hb)$\approx$20 $\AA$ and a significantly redder colour of 
$g-i\approx -0.4$ mag.
From Fig. \ref{g-i-evol}, we estimate the burst age of regions {\lvss a} and
{\lvss b} to be $\la$5 Myr and $\sim$100 Myr, respectively.
Both with regard to its absolute $g$ magnitude (--15.9 mag) and 
exponential scale length ($\alpha=0.19$ kpc), 
the LSB host of \object{J1044+0353} is typical of BCDs.
Its blue mean $g-i$ colour of $\sim$0 mag strongly suggests a predominantly 
young stellar population with an age between 0.1 and 1 Gyr for \sfha\ and \sfhb, respectively. 
\object{J1044+0353} therefore qualifies as a promising young XBCD candidate.
On the other hand, it cannot be ruled out from the available data that 
the blue $g-i$ colours of its host 
are partly due to extended ionized gas emission associated with the SF region {\lvss a}. 

Our derived oxygen abundance for region {\lvss a},
\oh=7.44$\pm$0.01, is close to the values of 
7.48$\pm$0.01 and 7.46$\pm$0.03 inferred from SDSS spectra 
by \citet{K03} and \citet{Izotov2006-SDSS}.
For the fainter region {\lvss b}, not studied previously,
we derive a comparably low oxygen abundance of \oh=7.45$\pm$0.09.

\smallskip
% =================================================
\object{J1201+0211}:  
% =================================================
Star-forming activities in this system are confined to within a single
compact (\P25=0.14 kpc) region at the southeastern tip of an elongated
irregular LSB host. This source contributes less than 10\% of the $g$ band
emission of the galaxy and has properties characteristic of a
very young stellar population ($g-i \approx$--0.7 mag, EW(\hb)=248 $\AA$).
Our oxygen abundance determination for this \h2\ region, 7.49$\pm$0.02, 
is consistent with previous determinations based on SDSS spectra
(7.55$\pm$0.03, Kniazev et al., 2003; 7.51$\pm$0.03, Izotov et al., 2006a).

Image stacking reveals a chain of faint ($m_g\geq$26 mag) compact sources 
discernible out to $\sim$1.4 kpc northwest of the H{\sc ii} region,
most notably regions labeled nw1 and nw2 in Fig. \ref{colour_maps}c. 
The irregular host galaxy of \object{J1201+0211} is relatively blue
($g-i=0.33\pm 0.1$ for \rr$\geq$4\farcs 6), suggesting an age of between 
1 and $\sim$2.3 Gyr for \sfha\ and \sfhb, respectively.

\smallskip
% =================================================
\object{J1414-0208}:  
% =================================================
This system bears morphological resemblance to the XBCD \object{2dF\ 115901}
(P06), however, in contrast to the latter, 
it shows mild SF activity. 
This is evident from its SBPs (Fig. \ref{SBPs}), which do not display
a central luminosity enhancement above the underlying host galaxy
and by its relatively faint nebular emission [EW(\hb)=48 $\AA$].
Its irregular LSB host has a blue mean $g-i$ 
colour of 0.23$\pm$0.1 mag, which suggests a dominant stellar population with
an age $<$2 Gyr (\sfhb).  
This system has not been studied previously. 
Its oxygen abundance of \oh=7.28$\pm$0.32, although relatively uncertain, 
places it within the XBCD metallicity range.
From SDSS spectra, we derive a comparably low oxygen abundance of \oh=7.35$\pm$0.26.

\smallskip
% =================================================
\object{J2230-0006}: 
% =================================================
With an exponential scale length of $\alpha=0.17$ kpc, \object{J2230-0006} 
is the most compact galaxy in our sample. Its LSB host shows slight deviations
from axis symmetry and a relatively blue $g-i$ colour of $\sim$0.3 mag.
SF activities are mainly present in the southern part of this XBCD
(cf. Fig. \ref{colour_maps}) and contribute about 20\% of its total
$g$ emission. The blue colour of its SF region (--0.6 mag within a
4\arcsec$\times$4\arcsec\ box) and its amble nebular emission 
[EW(\hb)=133 $\AA$]
suggest a young starburst age. 
The oxygen abundance of \object{J2230-0006} was determined to be
\oh=7.64$\pm$0.02, in good agreement with the value of \oh=7.66$\pm$0.04 by \citet{ITG2007}.

\smallskip
% =================================================
\object{J2302+0049}: 
% =================================================
This system contains two high surface brightness regions separated 
by 3\farcs4 ($\sim$2.2 kpc), which differ in their luminosity by a factor 
of approximately 5.
The brighter region {\lvss a} shows blue colours ($g-i=-0.78$ mag 
within a 2\arcsec$\times$2\arcsec\ box) and copious 
nebular emission (EW(\hb)=212 $\AA$, EW(\o5007)=1145 $\AA$) indicating
a substantial population of hot massive stars.
The oxygen abundance in this \h2\ region was determined from the 3.6m spectra 
to be \oh=7.63$\pm$0.02, close to the value of \oh=7.65$\pm$0.04 that we derive 
from SDSS spectra following \citet{Izotov2006-SDSS}.
The fainter region {\lvss b} shows no
emission lines and the signal-to-noise ratio of the available spectra does
not permit us to infer its redshift from stellar absorption lines.

The relatively large distance and intrinsic compactness of
\object{J2302+0049} prevent a detailed study of the properties of its
host galaxy, which on the available images is marginally more extended than the SF component. 
If primarily due to stellar emission, its mean colour of $\sim$0.1 mag, 
is consistent with a young evolutionary status with an age between 0.25 and
1.4 Gyr for \sfha\ and \sfhb, respectively.

% =================================================
\section{Discussion \label{discussion}}
% =================================================
Including the seven galaxies studied here, 
the total number of XBCDs has increased to $\sim$35.
It is therefore timely to review their general photometric and 
morphological properties, in order to explore 
possible trends and better coordinate future searches 
for these systems in the nearby universe and at higher redshift.

%%%%%%%%%%%%%%%%%%%%%%%%%%%%%%%%%%%%%%%%%%%%%%%%
%    Fig. 7 (structural properties) aa10028f33.ps
%%%%%%%%%%%%%%%%%%%%%%%%%%%%%%%%%%%%%%%%%%%%%%%%
\begin{figure*}
\hspace*{0.0cm}\psfig{figure=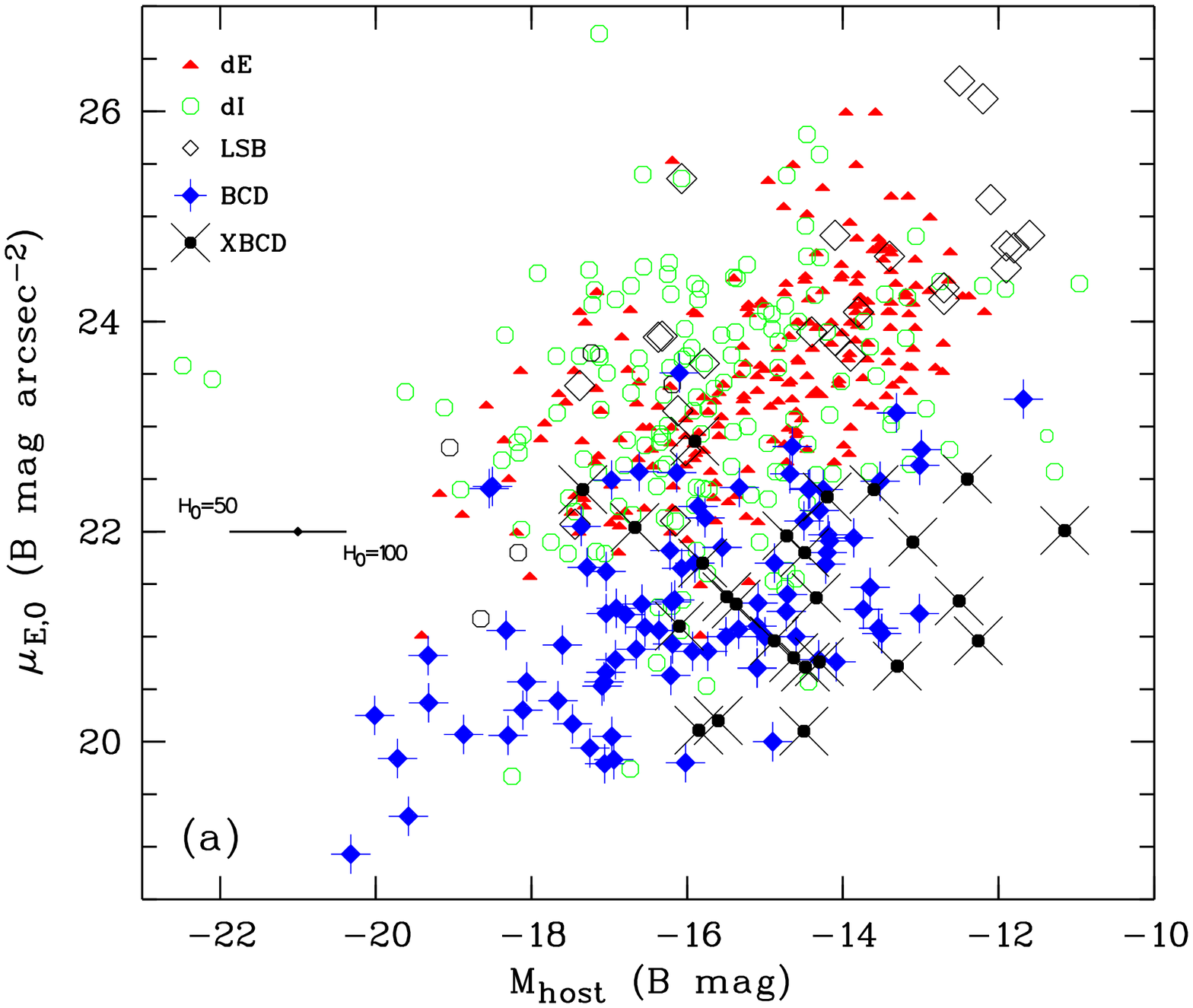,angle=-90.0,width=11.5cm,clip=}
\hspace*{-2.2cm}\psfig{figure=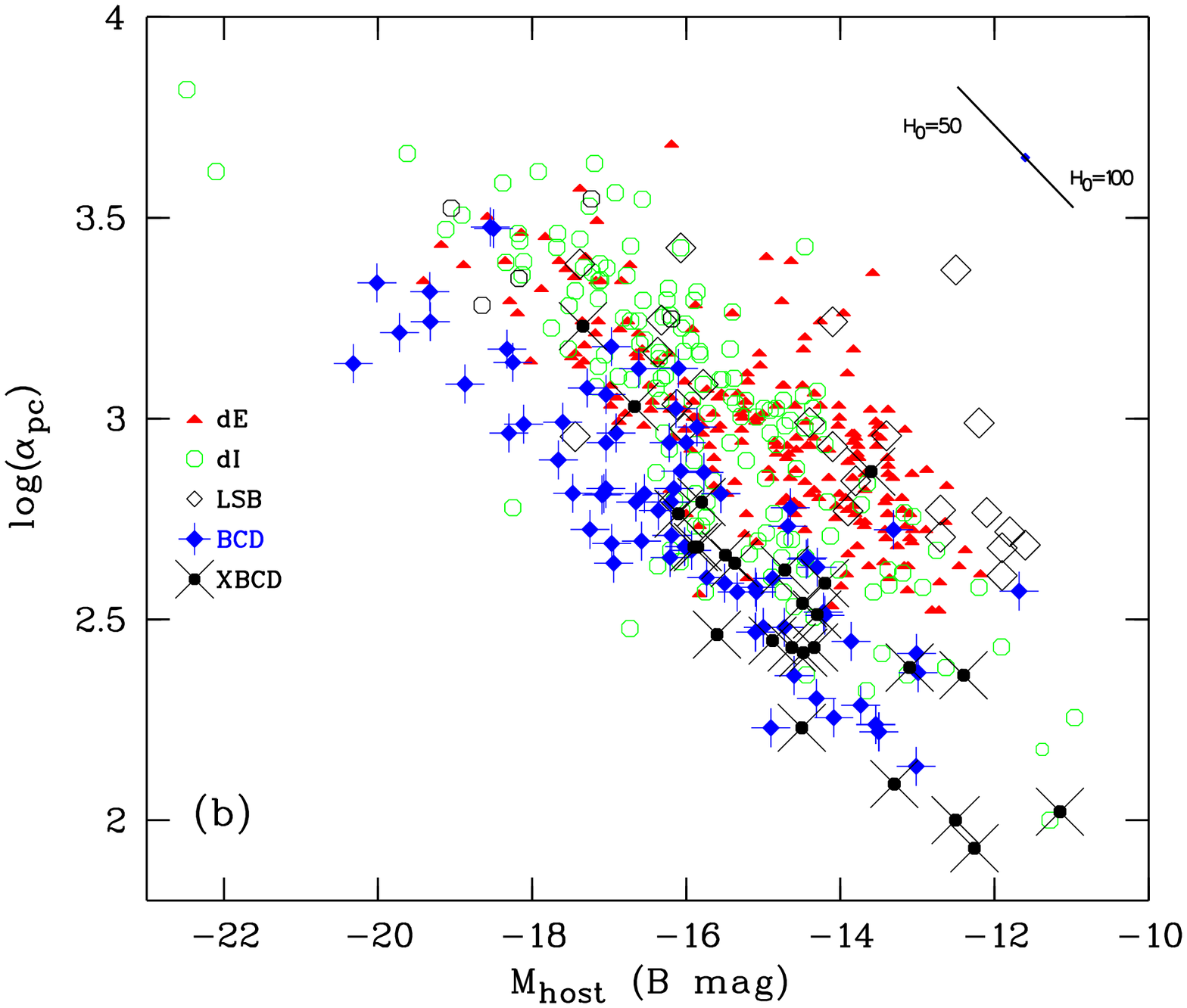,angle=-90.0,width=11.5cm,clip=}
\caption{Comparison of the structural properties of the host galaxy of XBCDs,
  BCDs, dIs, dwarf ellipticals (dEs) and Low-Surface Brightness (LSB)
  galaxies. 
Data for iE/nE BCDs are compiled from \citet{Cairos2001a}, \citet{drinkwater91},
\citet{Marlowe1997}, \citet{PapaderosFricke1998a}, \citet{Noeske2000-cometary},
  P96a, \citet{Papaderos1998-PhD} and P02.
Data for other types of dwarf galaxies are taken from
  \citet{BinggeliCameron91}, \citet{BinggeliCameron93}, \citet{Bothun91}, 
\citet{Caldwell87}, \citet{Carignan89}, \citet{Hopp91}, \citet{PT1996}, 
\citet{Vigroux86} and \citet{vanZee2000}.
Photometric quantities in the SDSS $g$ and $V$ band were transformed into
  $B$ band assuming $B - g = 0.3$ mag and $B - V = 0.5$ mag (see discussion in
  Sect. \ref{photo2} and in P06).
The lines show the shift of the data points caused by a change 
of the Hubble constant from 75 to 50 and 100 km s$^{-1}$ Mpc$^{-1}$.
{\bf (a)} Central surface brightness $\mu _{\rm E,0}$ vs. 
absolute $B$ magnitude $M_{\rm host}$ of the LSB component.
{\bf (b)} Logarithm of the exponential scale length $\alpha $ in pc 
vs. $M_{\rm host}$.
Data for XBCDs are compiled from
\citet{Papaderos1998-SBS0335}, \citet{Papaderos1999-Tol65}, \citet{Kniazev2000-HS0822},
\citet{Guseva2001-SBS0940}, \citet{Fricke2001-T1214}, P02,
\citet{Guseva2003-SBS1129}, \citet{Guseva2003-HS1442},
\citet{Guseva2003-SBS1415}, 
P06 and \citet{Pustilnik2005-DDO68}.
Five further systems with an oxygen abundance slightly above \oh=7.6 are also
included: \object{2dF 169299, UM 570, UM 559} and \object{2dF 84585}
[\oh=7.68, 7.71, 7.72, 7.66, respectively; P06] and
\object{Pox 186} \citep[\oh=7.74, ][]{Guseva2004-Pox186}.
}
\label{structure}
\end{figure*}
%%%%%%%%%%%%%%%%%%%%%%%%%%%%%%%%%%%%%%%%%%%%%%%%%
 
A first important conclusion from the present and previous studies 
is that, similar to other dwarf galaxies, the 
host galaxy of XBCDs can be approximated 
by an exponential fitting law in its outer parts, i.e. for 
galactocentric radii $3\,\alpha \la $\rr$\la 6\,\alpha$.
Additionally, several XBCDs display extended central flat cores 
in their host galaxies, reaching in some cases out to \rr$\sim$3$\alpha$. 
Centrally flattened exponential profiles have also been observed
in a sizeable fraction of early and late-type dwarfs spanning a wide 
range in absolute magnitude 
\citep[][P96a]{BinggeliCameron91,Noeske2003-NIR}.
The XBCD -- BCD connection is further supported by the fact 
that both object classes populate roughly the same locus in the $\mu_{\rm E,0}$ 
versus $M_{\rm host}$ and log($\alpha$) versus $M_{\rm host}$ parameter space 
(Fig. \ref{structure}) and are comparable in their effective radius (P06). 

These lines of evidence suggest that XBCDs do not represent peculiar cases 
of dwarf galaxy evolution, reflected in strongly distinct structural
properties (e.g. an abnormally diffuse or an ultra-compact LSB host), 
but that they share similar structural properties and are therefore likely 
also to lie on a common evolutionary track with the main population of 
more metal-rich BCDs with \oh$\ga$8.
Of course, this conclusion may not be free of selection biases, given 
that XBCDs are mostly detected by their high EW(\hb), blue colours, 
and compactness. It is therefore conceivable that a substantial population 
of relatively quiescent and more diffuse metal-poor SF
dwarfs remains strongly under-represented in current XBCD surveys.
The irregular SF dwarf \object{J0113+0052}, discovered by
\citet{Izotov2006-2XBCD}, may be regarded as an example of this kind. 
Additionally, it might be questioned that optical spectroscopic searches for XBCDs, 
based on forbidden line measurements in \h2\ regions, can uncover systems with 
a gas-phase metallicity significantly lower than 
\oh$\sim$7.0 \cite[$\sim$\zsun/60, 
adopting a solar abundance of 8.76,][]{Caffau2008}, 
thus better constrain the minimum level of 
chemical enrichment in the warm ISM of SF galaxies.
This is because, below some abundance level and ionization parameter, oxygen forbidden lines 
become too weak to be measurable.
For example, calculations with the photoionization code CLOUDY \citep{Ferland1998_CLOUDY90}
show that, at a metallicity \oh=7.0, electron density $N_{\rm e}$=100
cm$^{-3}$ and ionization parameter $U$=10$^{-3}$, the oxygen 
[O {\sc iii}]$\lambda$4363 and [O {\sc iii}]$\lambda$5007 line intensities
decrease to $<$1\% and 38\% of that of the H$\beta$ line, respectively, 
a determination of the electron temperature based on the [O {\sc iii}]$\lambda$4363
line becomes therefore practically impossible.

\smallskip
It is important to note that the detection of a \emph{stellar} LSB host in all known XBCDs, including
\object{SBS 0335-052E} \citep[][P98]{Izotov1997-SBS0335E,Thuan1997-SBS0335E} 
and \object{I Zw 18} \citep[][P02]{Izotov2001-IZw18,Papaderos2001-IZw18} 
suggests that none of these systems are currently forming \emph{in situ}
its first generation of stars.
This is also indicated by spatially resolved evolutionary synthesis studies 
\citep{Izotov1997-SBS0335E,Vanzi2000-SBS0335E,Guseva2001-SBS0940,Guseva2003-SBS1415,Hunt2003-IZw18}
and colour-magnitude diagram analyses of selected systems 
\citep[e.g.,][]{IzotovThuan2004-IZw18,IzotovThuan2004-UGC4483,OstlinMouhcine2005-IZw18,Aloisi2007-IZw18}.

On the other hand, the existing work consistently suggests that the host galaxies 
of XBCDs are less evolved than those of BCDs.
This is indicated by deep surface photometry studies that reveal 
uniformly blue colours in the XBCD hosts, with a
$V-I$ and $g-i$ index in the range between
$\sim$0.1 and $\la$0.5 mag 
\citep[e.g., P02;][and references therein]{Guseva2003-SBS1415} 
and $<$0.4 mag (this paper), respectively.
By contrast, the LSB hosts of iE/nE BCDs, representing $\sim$90\% of the BCD
population (LT86) show typically red ($\sim$1 mag) $B-R$ and $V-I$ colours 
\citep[P96b;][]{Cairos2001b,GildePazMadore2005}.

Another important, largely overlooked aspect concerns the morphology of XBCDs.
The hosts of these systems reveal in their majority conspicuous deviations
from axis-symmetry suggesting a little degree of dynamical relaxation.
By this, they again significantly differ from the \emph{bona fide} old, more metal-rich 
iE/nE BCDs whose defining property is a smooth \emph{elliptical} LSB host galaxy. 

%%%%%%%%%%%%%%%%%%%%%%%%%%%%%%%%%%%%%%%%%%%%%%%%
%    Fig. 8 (morphology)
%%%%%%%%%%%%%%%%%%%%%%%%%%%%%%%%%%%%%%%%%%%%%%%%
\begin{figure*}
\hspace*{0.7cm}\psfig{figure=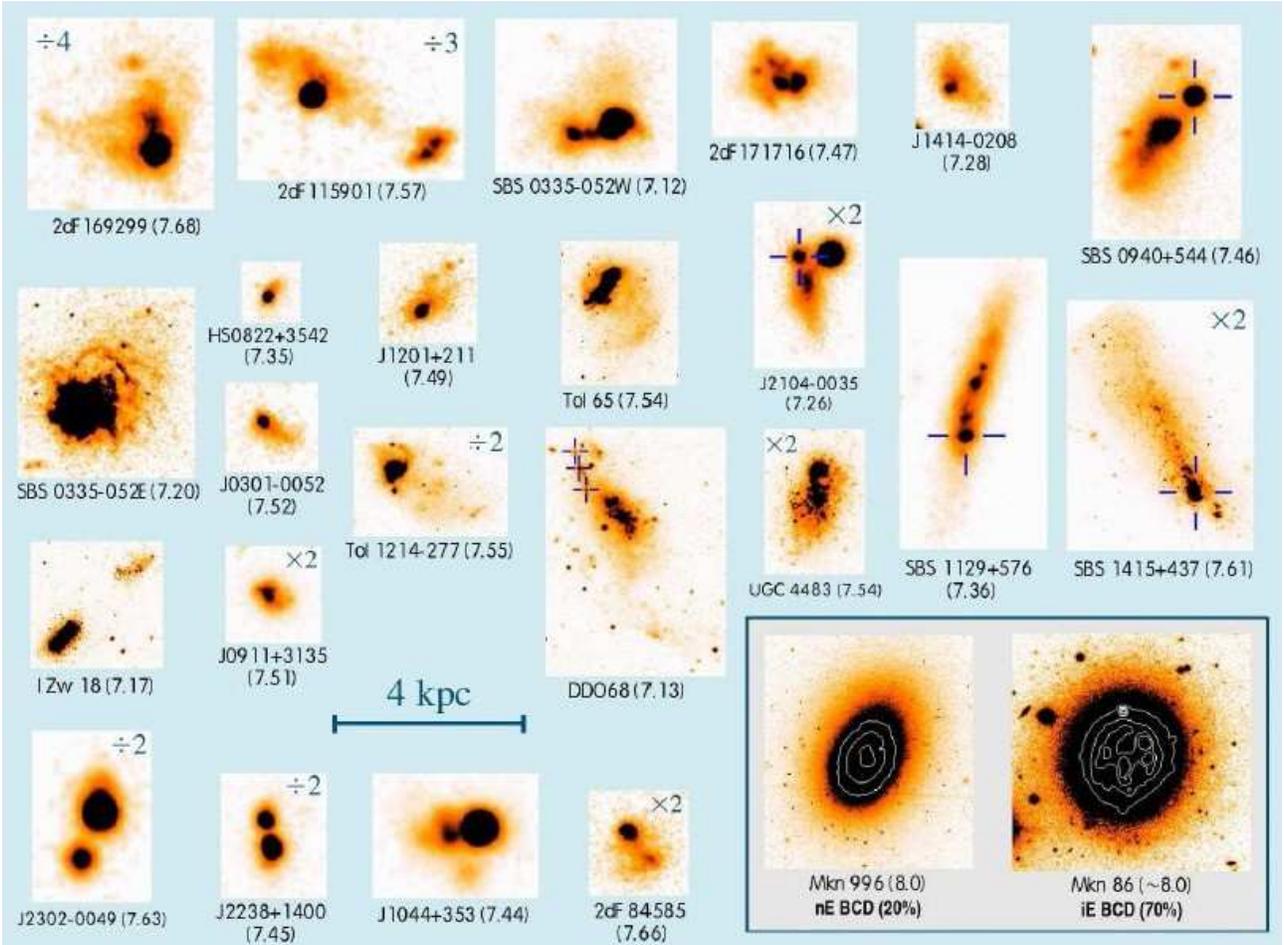,angle=0.0,width=17.0cm,clip=}
\caption{
 Comparison of systems in the range of XBCD metallicity with the main
  class of the more metal-rich, \emph{bona fide} old BCDs
  \object{Mkn 996} and \object{Mkn 86}. These two systems may be regarded as 
  prototypical examples of the dominant nE/iE morphological BCD type (cf. LT86).
  All galaxy images were scaled to a common distance to better facilitate 
  a comparison. Those images that are displayed
  magnified or downsized for the sake of better visibility are labeled with
 the respective scaling factor. Off-center low-metallicity \h2\ regions are
 marked with crosses for clarity. Within brackets we indicate the oxygen abundance
 of the lowest-metallicity \h2\ region in each XBCD. Metallicities are taken
 from the literature sources given in pages \pageref{irr_xbcds1} and \pageref{irr_xbcds2}.
 The images are from P06 (\object{2dF 169299, 2dF 115901, 2dF 171718, 2dF 84585}),
P98 (\object{SBS 0335-052E}, see also Thuan et
 al. 1997, HST program 5408), 
\citet[][\object{SBS 0335-052W}]{Papaderos2006-IAU},
\citet[][\object{SBS 0940+544}]{Guseva2001-SBS0940},
\citet[][\object{SBS 1129+576}]{Guseva2003-SBS1129},
\citet[][\object{SBS 1415+437}, HST program 5408]{Guseva2003-SBS1415},
\citet[][\object{I Zw 18}, $R$ band HST WFPC2 exposure with
  the \ha\ emission subtracted out, HST program 5309]{Papaderos2001-IZw18},
the HST archive (\object{Tol 65, Tol 1214-277, Mkn 996}; HST programs
5408 and 6678),
Papaderos \& Noeske (2008, in preparation, \object{HS 0822+3542, UGC 4483, Mkn 86}),
and from the SDSS (\object{J1414-0208, J1044+353, J2302-0049, J1201+211, J2104-0035, J0301-0052,
  J0911+3135, DDO68, J2238+1400}).
}
\label{XBCD-morph}
\end{figure*}

%%%%%%%%%%%%%%%%%%%%%%%%%%%%%%%%%%%%%%%%%%%%%%%%%

Faint, non-axisymmetric distortions are present in roughly one half of the 
XBCDs studied here. 
The irregular outer morphology of those systems is certainly 
not due to insufficient detection of their underlying galaxy host.
This is because, as is evident from Fig. \ref{colour_maps}, SDSS images 
allow us to interpolate contours down to a $g$ band surface brightness $\mu_g \geq
25.5$ \sbb, corresponding to $\mu_B \approx 25.8$ \sbb.
At such intensity levels the elliptical host of evolved BCDs 
dominates the light \citep[P96b, ][P02]{Cairos2001b}
and should have been detected if it were present. 
Furthermore, with the possible exception of \object{J1201+0211},
stacked $gri$ images do not reveal tidal features in any of
our sample SDSS galaxies, ruling out \emph{strong} gravitational interactions
or galaxy merging as the origin of the observed morphological distortions.
Widespread SF activity in the LSB host can also be excluded 
from $g-i$ colour maps and long-slit spectra.

Likewise, most of the XBCDs investigated previously on the basis of deep surface
photometry are as well characterised by irregular host galaxies.
\label{irr_xbcds1}
Such examples are 
\object{SBS 0335-052W} \citep[\oh=7.12, ][see P98 and Papaderos et al. 2006c 
for photometry]{Izotov2005-SBS0335W},
\object{Tol 65} \citep[\oh=7.54, ][see Papaderos et al. 1999 for photometry]{Izotov2004-T1214-T65},
\object{SBS 1415+437} \citep[\oh=7.6,][]{Thuan1999-SBS1415,Guseva2003-SBS1415},
\object{Tol 1214-277} \citep[\oh=7.55, ][see Fricke et al. 2001 for photometry]{Izotov2004-T1214-T65},
\object{I Zw 18} \citep[\oh=7.17, ][see Papaderos et al. 2001 and P02 for photometry]{Izotov1997-IZw18},
\object{SBS 0940+544} \citep[\oh=7.46, ][]{Guseva2001-SBS0940},
\object{SBS 1129+576} \citep[\oh=7.36, ][]{Guseva2003-SBS1129},
\object{HS 0837+4717} \citep[\oh=7.64,][]{Pustilnik2004-HS0837},
\object{DDO 68} \citep[\oh=7.21\dots 7.13, ][respectively]{Pustilnik2005-DDO68,IzotovThuan2007-MMT},
\object{HS 2134+0400} \citep[\oh=7.44,][]{Pustilnik2006-HS2134,Guseva2007-BJ},
\object{J2104-0035} and \object{J0113+0052} 
\citep[\oh$\approx$7.2 and 7.26, respectively,][]{Izotov2006-2XBCD},
\object{2dF 171716, 2dF 115901} (\oh=7.5, 7.57, respectively; P06)
and 
\object{J0301-0052}, \object{J0911+3135}, \object{J2238+1400}
\citep[\oh=7.52, 7.51 and 7.56, respectively, ][]{IzotovThuan2007-MMT}.
This is also the case for several systems with an oxygen abundance slightly
above \oh=7.6, such as \object{2dF 84585} and \object{2dF 169299} 
(\oh=7.66 and 7.68, respectively; P06).

\smallskip
With regard to the morphological properties of the star-forming component of XBCDs 
(e.g. multiplicity, luminosity distribution, and the 
degree of the confinement of SF regions to the center of the XBCD host),
these have never been studied in a systematic manner before, and a comparison 
with iE/nE BCDs would therefore be premature at this point.
However, two trends are apparent. First, in a significant fraction of XBCDs, 
star-forming activities are not strongly confined to the geometrical center 
of the host galaxy and, second, the global SF process in several of these
systems appears to be largely driven by propagation.

Figure \ref{XBCD-morph} contains several examples of XBCDs  
with off-center SF activity, including \object{2dF 115901}, 
\object{SBS 0335-052W}, \object{J1414-0208}, and \object{J0911+3135}.
The most impressive instances of off-center SF activity among XBCDs are
\emph{cometary} systems (iI,C type in the BCD classification
scheme devised by LT86).
These objects contain a luminous SF region 
at the one tip of an elongated blue and irregular host galaxy with a 
gradually decreasing surface brightness towards its antipodal end. 
The hypothesis that these systems are edge-on disks with a dominant 
SF complex in their outermost periphery can be dismissed on statistical 
grounds. 
Some iI,C XBCDs (for example, \object{SBS 1415+437} and \object{SBS 0940+544}) 
display signatures of low-level ongoing or recent star formation 
along their major axis or colour gradients suggesting propagation of SF activities 
from the far-end side of their elongated host galaxy towards the young, dominant SF region
\citep[see e.g. P98,][]{Guseva2001-SBS0940,Guseva2003-SBS1415}.

That the formation of the stellar component in a XBCD may largely 
be driven by SF propagation has been suggested from an analysis of
\label{irr_xbcds2}
\object{SBS 0335-052E} 
\cite[\oh=7.3\dots
  7.2,][respectively]{Izotov1997-SBS0335E,Papaderos2006-SBS0335}, 
for which P98 have estimated a SF propagation velocity of $\sim$20 km/sec, 
of the order of the sound speed in the warm ISM.
It is unclear for how long this process can continue; however, 
as long as gas supply of the appropriate temperature and density 
is available ahead of the SF front, no self-limiting mechanism should exist. 
Propagating star formation with a constant speed $u$ over a period  
$\tau$ may naturally lead to a cometary morphology on a linear scale 
$l \sim$ 10 kpc $\times (u/10\,\,{\rm km/s}) \times (\tau/1\,\,{\rm Gyr})$.
As evident from Fig. \ref{XBCD-morph}, $l$ is of the order of
the projected major-axis of several cometary XBCDs. 

We note that the iI,C morphology is not uniquely observed in XBCDs and that
examples of this kind also exist among BCDs of higher metallicity
\citep{Noeske2000-cometary,Cairos2001a,Noeske2003-NIR,GildePaz2003}.
The essential trend, however, is that whereas systems with cometary morphology
or strongly off-center SF activity comprise less than 10\% of the BCD
population (LT86), they apparently dominate the XBCD population.
In this context, it is worth pointing out that studies of iI,C BCDs 
in the metallicity range between $\approx$7.8 and 8.0 by \cite{Noeske2000-cometary}
suggest that these systems are younger ($\la$4 Gyr) 
than the main class of iE/nE BCDs, they are therefore likely to represent 
intermediate stages of BCD evolution. 
This conjecture is consistent with the high incidence of cometary 
systems among young XBCD candidates discussed here.

It is unclear whether or not cometary morphology is linked to galaxy interactions, 
as neither observations nor numerical simulations presently provide tight
constraints in this respect.
However, it is known that a significant fraction ($>$30\%) of BCDs are not
truly isolated but have optically faint nearby companions 
\citep{Noeske2001,Pustilnik2001-env}.
Dwarf galaxy encounters with a wide range of impact parameters are therefore 
likely to play an important role in BCD evolution. 
The frequency of strong collisions and, eventually, subsequent merging of BCD
progenitors is still not constrained well. However, several notable 
examples (iI,M BCDs in the classification scheme of LT86) exist 
in the samples of \cite{Cairos2001a} and \cite{GildePaz2003}, and there 
is growing evidence that the major fraction of intrinsically luminous ($M_B<$--18 mag) 
Blue Compact Galaxies are of merger origin \citep{Ostlin2001-BCGII}.
In addition, the morphology of some XBCDs (for example, \object{2dF 169299}, 
\object{2dF 115901}, and \object{2dF 171716}; cf. Fig. \ref{XBCD-morph}) is 
consistent with the merger interpretation.
The importance of gravitational interactions to BCD evolution 
is indicated further by radio interferometry that reveals HI clouds 
in the close vicinity ($\la$100 kpc) of numerous systems \citep{Taylor1993,THL2004}.

The hypothesis that \emph{strong} gravitational interactions or galaxy merging
are the origin of cometary morphology does not however appear to be tenable.
If cometary BCDs were indeed forming by galaxy merging, one would  
expect tidal features to protrude far beyond their Holmberg radius and be 
readily detectable at surface brightness levels of $\mu\sim$24 $B$ \sbb,
in a similar way to merging disk galaxies.
Since the visibility timescale of tidally ejected stellar and gaseous matter 
is long \cite[$\sim$1 Gyr;][and references therein]{HibbardMihos1995},
of the order of the 
luminosity-weighted age of the XBCD galaxy host, these features should be
almost ubiquitous in iI,C systems.
The probability that both merging counterparts are metal poor
and retain their gas-phase metallicity of a level  
\oh$\la$7.6, even after a strong, merging-induced starburst, is also low.

A more viable interpretation involves \emph{weak} interactions with 
low-mass stellar or gaseous companions. These may have a twofold effect, 
leading to a bar-like gas distribution and triggering SF activities
that, by propagation along the direction of maximum gas density, 
could subsequently produce a cometary BCD morphology. 
Alternatively, a propagating shock wave induced by gas-cloud infall 
onto a quiescent late-type dwarf might generate a similar star formation
pattern.

In summary, the hypothesis that SF propagation is the main process 
driving the formation of cometary XBCDs is not in conflict 
with the hypothesis that the evolution of these systems is largely
influenced by interactions.
However, lacking theoretical guidence and robust statistics on the 
gas distribution and kinematics of these systems, the 
possible role of interactions cannot be reliably assessed.
However, numerical simulations of increasing sophistication continue to
reproduce important properties of SF dwarfs 
\cite[e.g.][]{Noguchi2001,Recchi2002,Pelupessy2004,Pelupessy2006,
Hensler2004,Bekki2008},
and promise to provide key insights into the star formation history and
morphological evolution of cometary dwarfs in the near future.
Dedicated observational studies of these extremely metal-poor cometary galaxies 
will also hopefully provide important constraints and 
incentives for theoretical work.

\smallskip
As mentioned above, a compelling argument that XBCDs have undergone 
previous evolution derives from the detection of an extended \emph{stellar} 
host galaxy.
The blue, luminosity-weighted colours of this component, interpreted 
in the framework of simple SFH parametrisations (e.g., {\lvss SFH2}) suggest 
that XBCDs form a heterogeneous class of 
predominantly cosmologically young objects, with examples among them which 
have formed most of their stellar mass in the past few $10^8$ yrs
and are possibly still experiencing the major phase of their dynamical assembly 
(e.g. \object{SBS 0335-052E}, \object{I Zw 18}) 
to moderately evolved cometary systems 
that likely formed most of their stellar mass within the last $\sim$2 Gyr
(\object{Tol 65}, \object{Tol 1214-277}, \object{SBS 1415+437}, 
\object{SBS 0940+544}, \object{J1044+0353}).
This conclusion is supported further by detailed evolutionary synthesis models
\cite[e.g.][]{Izotov2001-IZw18,Guseva2001-SBS0940,Guseva2003-SBS1129,Guseva2003-SBS1415}
involving more complex SFHs and which aim to account self-consistently for
a variety of observables, such as the EWs of Balmer
emission and absorption lines, the slope of the spectral energy
distribution (SED), and spatially resolved colours. 

In the case of the XBCD \object{I Zw 18}, CMD studies based on HST ACS images 
do not converge into a broadly accepted interpretation 
about the age of its stellar component, yielding 
values between $\leq$0.5 Gyr \citep{IzotovThuan2004-IZw18} and $>$1 Gyr
\citep{Aloisi2007-IZw18}.
This might be partly due to the specific properties of this system.
\cite{Papaderos2002-IZw18} showed that 
the exponential LSB host of \object{I Zw 18}, which was previously thought 
to be dominated by stellar emission, is due entirely to extended, patchy ionized 
gas emission, therefore extreme caution is required when using CMDs and surface photometry 
to place constraints on the formation history of its stellar component.
The \emph{stellar} component of \object{I Zw 18} is by a factor of 
approximately 2 more compact that the ionized gas halo and, 
in contrast to the main class of 
evolved BCDs, has uniformly blue colours down to a surface brightness level 
of $\mu \sim$ 26 $B$ \sbb. 
Additionally, as found by \cite{Cannon2002-IZw18}, \object{I Zw 18} shows a highly
inhomogeneous extinction pattern, a fact that further complicates CMD analyses.
The nature of point sources detected in the presense of dominant ionized gas emission 
in \object{I Zw 18}, at a distance of 18.2 Mpc \citep{Aloisi2007-IZw18}, 
which corresponds to a projected area of $\sim$20 pc$^2$ per HST ACS pixel,
is undoubtedly an important question in XBCD research.

The formation history of the XBCD host is clearly a crucial and outstanding issue.
XBCDs may all contain a faint substrate of stars of cosmological age 
(see discussion in P98), quite similar to the ancient 
metal-poor stellar population in Local Group dwarf spheroidals \citep{GrebelGallagher2004}.
As pointed out in P98, this putative ancient
stellar background cannot be entirely ruled out on spectrophotometric grounds, 
since its effect on the observed SED might be barely detectable. 
It is only possible to infer an upper bound to the mass fraction 
of this hypothetical old stellar population that, by necessity, 
is tied to simplifying assumptions about the SFH and intrinsic extinction.
Current upper limits for a few thoroughly analysed XBCDs
range between $\sim$15\% and $<$50\% \citep{Vanzi2000-SBS0335E,Guseva2001-SBS0940,
Guseva2003-SBS1415,Hunt2003-IZw18,Pustilnik2004-SBS0335}.

Arguably, a low mass fraction of ancient stars possibly present in XBCDs 
do not rule out the young evolutionary status of these systems.
This would only be the case if the formation epoch of those first stars were 
coeval with the dominant phase of XBCD formation, and XBCDs/BCDs were invariably
forming in a single, short (1--3 Gyr) SF episode that converted most of 
their gaseous reservoir into stars. This interpretation is
untenable, however, inter alia, because of the gas-richness and 
recurrent starburst activity of these systems 
\citep[see e.g.][]{Thuan1991,MasHesseKunth1999}.
As a matter of fact, dwarf galaxies of a given baryonic mass may have followed
quite diverse evolutionary pathways.
This is manifestated in e.g. the Local Group galaxies for which we observe an 
impressive manyfold of SFHs and morphologies, 
comprising both ancient metal-poor dwarf spheroidals and metal-rich dEs 
\citep[e.g. \object{NGC 185, NGC 205}][see also Mateo 1998]{Dolphin2005} 
as well as late-type irregulars that underwent the dominant phase
of their build-up only $\sim$1 Gyr ago 
\citep[e.g. \object{Sextans A}][]{Dolphin2003-SexA}.
The inherent variety of SFHs in dwarf galaxies is further enhanced by the role of
the environment, which is recognised to be an important factor in galaxy evolution 
\citep{BO1984,Dressler1980,Dressler1997,Poggianti1997,Pustilnik2001-env}.
For example, it is well established that the EW(\hb) of SF dwarfs increases towards the
periphery of galaxy clusters \citep{Vilchez1995} and that 
XBCDs preferentially populate the extreme field \citep{Pustilnik2001-env}.
These findings are consistent with the idea that the formation timescale
of relatively isolated late-type dwarfs is the longest, making low-density
regions promising sites to search for young XBCDs candidates.
In view of such considerations, the existence of a small number of 
unevolved XBCDs in the nearby universe is unsurprising, and a 
manifestation of the diversity in the SFHs of dwarf galaxies. 

% ================================ Summary
\section{Summary \label{summary}}
% ================================ 

We have presented spectroscopic observations with the 3.6m ESO telescope
of eight H {\sc ii} regions in seven emission-line galaxies, 
selected from the Data Release 4 (DR4) of the Sloan Digital Sky Survey
(SDSS) (six galaxies) and from the Six-Degree Field Galaxy Redshift Survey (6dFGRS) 
(one galaxy). 
The three brightest of these sources were first identified by \cite{K03, K04b} 
using SDSS data and their oxygen abundances were measured using SDSS spectra.

From our 3.6m long-slit data we determined the oxygen abundance of these systems
to be \oh$\la$7.6, which places them among the most metal-poor star-forming
(SF) galaxies ever discovered. 
Furthermore, we used imaging data to study the morphology and
surface brightness and colour distribution of the SDSS galaxies in our sample.
From decomposition of surface brightness profiles (SBPs), we infered a 
wide range of between $\sim$5\% and $\sim$50\% for the luminosity contribution 
of the SF component to the total $g$ band emission.
We also measured a wide range of equivalent widths for the  
strong nebular lines, varying between $\sim$20 $\AA$ and $\sim$320 $\AA$ 
for the H$\beta$ line and between $\sim$40 $\AA$ and $\sim$1400 $\AA$ for 
the [O{\sc iii}]$\lambda$5007 line.
The effect of copious ionized gas emission manifests itself
also in extraordinarily blue colours ($g-i\la -0.8$ mag) in the SF component
of several galaxies studied.

\smallskip
The present investigation of extremely metal-poor SF galaxies (hereafter
XBCDs) in conjunction with data from the literature leads to the following conclusions:

\begin{enumerate}
\item[i)] all XBCDs known possess a compact to moderately extended
  \emph{stellar} host galaxy underlying their SF component, thus they are not experiencing
the \emph{in situ} formation of their first stellar generation in a galaxy-wide starburst.

\item[ii)] XBCDs are with respect to the exponential outer slope and the structural
  properties of their host galaxies fairly comparable to the main type
  ($\sim$90\%)
  of the \emph{bona fide} old and more metal-rich iE/nE blue compact dwarf (BCD) galaxies.
  This suggests that XBCDs do not represent peculiar cases of dwarf galaxy
  evolution, reflected by their distinctive structural properties 
(e.g. abnormally diffuse or ultra-compact host galaxies), but that they 
probably follow a common evolutionary track with the main population of BCDs.

\item[iii)] 
The host galaxies of XBCDs are substantially bluer than the
hosts of iE/nE BCDs, suggesting that the dominant mass fraction 
of the stellar component in these systems formed
within the past $\sim$2 Gyrs. 
The possibility that XBCDs contain a low mass fraction of old ($\sim$10 Gyr) stars
cannot be dismissed.

\item[iv)] 
The host galaxies of XBCDs typically show conspicuous morphological 
distortions in their periphery, suggesting a little degree of dynamical relaxation.
In this property, they again markedly differ from the main class of the old, elliptical BCDs.
XBCDs are also different from typical BCDs in terms of their lower spatial confinement of 
SF activities to their geometrical center. 
Most notably, a large fraction of XBCDs show a \emph{cometary} morphology,
characterised by the presence of a dominant star-forming region at one tip of
  an elongated, blue stellar host. 
Surface brightness and colour gradients along that host galaxy can be plausibly 
accounted for by propagation of star-forming activities along its body.

\end{enumerate}

The overall conclusion emerging from the present study is that XBCDs 
form a heterogeneous class of predominantly cosmologically young objects, 
with examples among them which are still undergoing the major phase of their 
formation, to moderately evolved cometary systems that have formed the 
dominant mass fraction of their stellar component within the past $\sim$2 Gyrs.
XBCDs may therefore hold important insights into the early morphological and 
dynamical evolution of low-mass protogalaxies in the distant universe.

\begin{acknowledgements}
P.P. would like to thank Prof. A. Zensus and 
the Max-Planck-Institute for Radioastronomy in Bonn for their hospitality.
N.G.G. and Y.I.I. thank the hospitality of the Institute for Astrophysics
(G\"ottingen) and the support of the DFG grant No. 436 UKR 17/15/06.
P.P. would like to thank Gaspare Lo Curto, Lorenzo Monaco, Carlos La Fuente, 
Eduardo Matamoros and the whole ESO staff at the La Silla Observatory for 
their support. 
Y.I.I. acknowledges the partial financial support of NSF
grant AST 02-05785. 
All the authors acknowledge the work of the Sloan Digital Sky
Survey (SDSS) team.
Funding for the SDSS has been provided by the
Alfred P. Sloan Foundation, the Participating Institutions, the National
Aeronautics and Space Administration, the National Science Foundation, the
U.S. Department of Energy, the Japanese Monbukagakusho, and the Max Planck
Society. The SDSS Web site is http://www.sdss.org/.
We would like to thank the Anglo-Australian Observatory staff at
the UK Schmidt Telescope and the entire 6dFGS team for ensuring
the success of the 6dFGS. 
This research has made use
of the NASA/IPAC Extragalactic Database (NED) which is operated by the 
Jet Propulsion Laboratory, CALTECH, under contract
with the National Aeronautic and Space Administration. 
This publication makes use of data products from the Two Micron All Sky
Survey, which is a joint project of the University of Massachusetts
and the Infrared Processing and Analysis Center/California Institute
of Technology, funded by the National Aeronautics and Space
Administration and the National Science Foundation.
We are grateful to the anonymous referee for her/his helpful comments and suggestions.
\end{acknowledgements}

\end{document}